\documentclass[twocolumn,floatfix,showpacs,nofootinbib]{revtex4}
\usepackage{graphicx}
\usepackage{amsmath}
\begin{document}
\title{Andreev billiards}
\author{C. W. J. Beenakker}
\affiliation{Instituut-Lorentz, Universiteit Leiden, P.O. Box 9506, 2300 RA
Leiden, The Netherlands}
\date{July 2004}
\begin{abstract}
This is a review of recent advances in our understanding of how Andreev
reflection at a superconductor modifies the excitation spectrum of a quantum
dot. The emphasis is on two-dimensional impurity-free structures in which the
classical dynamics is chaotic. Such Andreev billiards differ in a fundamental
way from their non-superconducting counterparts. Most notably, the difference
between chaotic and integrable classical dynamics shows up already in the level
density, instead of only in the level--level correlations. A chaotic billiard
has a gap in the spectrum around the Fermi energy, while integrable billiards
have a linearly vanishing density of states. The excitation gap $E_{\rm gap}$
corresponds to a time scale $\hbar/E_{\rm gap}$ which is classical
($\hbar$-independent, equal to the mean time $\tau_{\rm dwell}$ between Andreev
reflections) if $\tau_{\rm dwell}$ is sufficiently large. There is a competing
quantum time scale, the Ehrenfest time $\tau_{E}$, which depends
logarithmically on $\hbar$. Two phenomenological theories provide a consistent
description of the $\tau_{E}$-dependence of the gap, given qualitatively by
$E_{\rm gap}\simeq\min(\hbar/\tau_{\rm dwell},\hbar/\tau_{E})$. The analytical
predictions have been tested by computer simulations but not yet
experimentally.
\end{abstract}
\pacs{74.45.+c,05.45.Mt,73.23.-b,74.78.Na}
\maketitle

\section{Introduction}
\label{intro}

\begin{figure}
\centerline{\includegraphics[width=8cm]{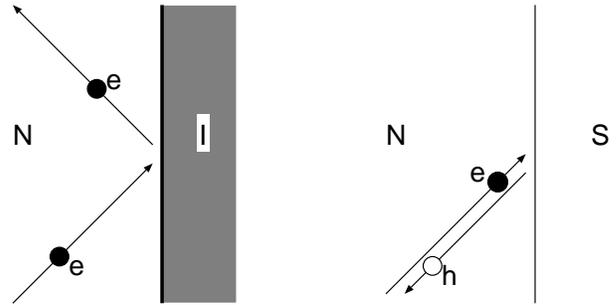}}
\caption{
Normal reflection by an insulator (I) versus Andreev reflection by a
superconductor (S) of an electron excitation in a normal metal (N) near the
Fermi energy $E_{F}$. Normal reflection (left) conserves charge but does not
conserve momentum. Andreev reflection (right) conserves momentum but does not
conserve charge: The electron (e) is reflected as a hole (h) with the same
momentum and opposite velocity. The missing charge of $2e$ is absorbed as a
Cooper pair by the superconducting condensate. The electron-hole symmetry is
exact at the Fermi level. If the electron is at a finite energy $E$ above
$E_{F}$, then the hole is at an energy $E$ below $E_{F}$. The energy difference
of $2E$ breaks the electron-hole symmetry. From Ref.\ \cite{Bee97}.
\label{reflection}
}
\end{figure}

Forty years ago, Andreev discovered a peculiar property of superconducting
mirrors \cite{And64}. As illustrated in Fig.\ \ref{reflection}, an electron
that tries to enter a superconductor coming from the Fermi level of a normal
metal is forced to retrace its path, as if time is reversed. Also the charge of
the particle is reversed, since the negatively charged electron is converted
into a positively charged hole. The velocity of a hole is opposite to its
momentum, so the superconducting mirror conserves the momentum of the reflected
particle. In contrast, reflection at an ordinary mirror (an insulator)
conserves charge but not momentum. The unusual scattering process at the
interface between a normal metal (N) and a superconductor (S) is called Andreev
reflection.

\begin{figure}
\centerline{\includegraphics[width=6cm]{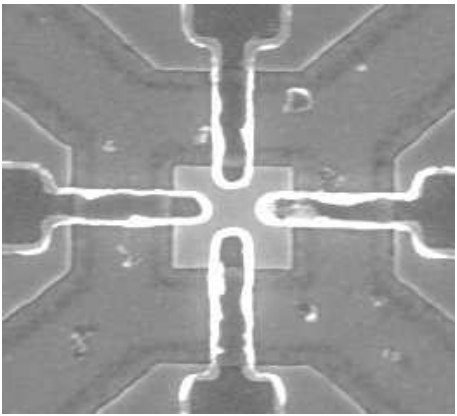}}
\caption{
Quantum dot (central square of dimensions $500\,{\rm nm}\times 500\,{\rm nm}$)
fabricated in a high-mobility InAs/AlSb heterostructure and contacted by four
superconducting Nb electrodes. Device made by A. T. Filip, Groningen University
(unpublished figure).
\label{andrei}
}
\end{figure}

Andreev reflection is the key concept needed to understand the properties of
nanostructures with NS interfaces \cite{Imr02}. Most of the research has
concentrated on transport properties of open structures, see Refs.\
\cite{Bee97,Wee97} for reviews. There experiment and theory have reached a
comparable level of maturity. In the present review we focus on spectral
properties of closed structures, such as the quantum dot with superconducting
contacts shown in Fig.\ \ref{andrei}. The theoretical understanding of these
systems, gained from the combination of analytical theory and computer
simulations, has reached the stage that a comprehensive review is called for
--- even though an experimental test of the theoretical predictions is still
lacking.

An impurity-free quantum dot in contact with a superconductor has been called
an ``Andreev billiard'' \cite{Kos95}.\footnote{Open structures containing an
antidot lattice have also been called ``Andreev billiards'' \cite{Ero02}, but
in this review we restrict ourselves to closed systems.}
The name is appropriate, and we will use it too, because it makes a connection
with the literature on quantum chaos \cite{Gut90,Haa01}. A billiard (in the
sense of a bounded two-dimensional region in which all scattering occurs at the
boundaries) is the simplest system in which to search for quantum mechanical
signatures of chaotic classical dynamics. That is the basic theme of the field
of quantum chaos. By introducing a superconducting segment in the boundary of a
billiard one has the possibility of unraveling the chaotic dynamics, so to say
by making time flow backwards. Andreev billiards therefore reveal features of
the chaotic dynamics that are obscured in their normal (non-superconducting)
counterparts.

The presence of even the smallest superconducting segment suppresses the
quantum mechanical level density at sufficiently low excitation energies. This
suppression may take the form of an excitation gap, at an energy $E_{\rm gap}$
well below the gap $\Delta$ in the bulk superconductor (hence the name
``minigap''). It may also take the form of a level density that vanishes
smoothly (typically linearly) upon approaching the Fermi level, without an
actual gap. The presence or absence of a gap is a quantum signature of chaos.
That is a fundamental difference between normal billiards and Andreev
billiards, since in a normal billiard the level density can not distinguish
chaotic from integrable classical dynamics. (It depends only on the area of the
billiard, not on its shape.)

A powerful technique to determine the spectrum of a chaotic system is
random-matrix theory (RMT) \cite{Bee97,Meh91,Guh98}. Although the appearance of
an excitation gap is a quantum mechanical effect, the corresponding time scale
$\hbar/E_{\rm gap}$ as it follows from RMT is a classical (meaning
$\hbar$-independent) quantity: It is the mean time $\tau_{\rm dwell}$ that an
electron or hole excitation dwells in the billiard between two subsequent
Andreev reflections. A major development of the last few years has been the
discovery of a competing quantum mechanical time scale
$\tau_{E}\propto|\ln\hbar|$. (The subscript E stands for Ehrenfest.) RMT breaks
down if $\tau_{E}\agt\tau_{\rm dwell}$ and a new theory is needed to determine
the excitation gap in this regime. Two different phenomenological approaches
have now reached a consistent description of the $\tau_{E}$-dependence of the
gap, although some disagreement remains.

The plan of this review is as follows. The next four sections contain
background material on Andreev reflection (Sec.\ \ref{Andreev}), on the minigap
in NS junctions (Sec.\ \ref{minigap}), on the scattering theory of Andreev
billiards (Sec.\ \ref{scatteringformu}), and on a stroboscopic model used in
computer simulations (Sec.\ \ref{strobomodel}). The regime of RMT (when
$\tau_{E}\ll\tau_{\rm dwell}$) is described in Sec.\ \ref{RMT} and the
quasiclassical regime (when $\tau_{E}\gg\tau_{\rm dwell}$) is described in
Sec.\ \ref{quasiclassics}. The crossover from $E_{\rm gap}\simeq\hbar/\tau_{\rm
dwell}$ to $E_{\rm gap}\simeq\hbar/\tau_{E}$ is the topic of Sec.\
\ref{clqcrossover}. We conclude in Sec.\ \ref{concl}.

\section{Andreev reflection}
\label{Andreev}

The quantum mechanical description of Andreev reflection starts from a pair of
Schr\"{o}dinger equations for electron and hole wave functions $u({\bf r})$ and
$v({\bf r})$, coupled by the pair potential $\Delta({\bf r})$. These socalled
Bogoliubov-De Gennes (BdG) equations \cite{DeG66} take the form
\begin{eqnarray}
&&{\cal H}_{\rm BG}\left(\begin{array}{c}u\\v\end{array}\right)=E
\left(\begin{array}{c}u\\v\end{array}\right),\label{BdG1}\\
&&{\cal H}_{\rm BG}=\left(\begin{array}{cc} H&{\Delta}({\bf r})\\
{\Delta}^{\ast}({\bf r})&-H^{\ast} \end{array}\right).\label{HBG}
\end{eqnarray}
The Hamiltonian $H=({\bf p}+e{\bf A})^{2}/2m+V-E_{F}$ is the single-electron
Hamiltonian in the field of a vector potential ${\bf A}({\bf r})$ and
electrostatic potential $V({\bf r})$. The excitation energy $E$ is measured
relative to the Fermi energy $E_{F}$. If $(u,v)$ is an eigenfunction with
eigenvalue $E$, then $(-v^{\ast},u^{\ast})$ is also an eigenfunction, with
eigenvalue $-E$. The complete set of eigenvalues thus lies
symmetrically around zero. The quasiparticle excitation spectrum consists of
all
positive $E$.

In a uniform system with $\Delta({\bf r})\equiv\Delta$, ${\bf A}({\bf r})\equiv
0$, $V({\bf r})\equiv 0$, the
solution of the BdG equations is
\begin{eqnarray} E&=&\left[
(\hbar^{2}k^{2}/2m-E_{F})^{2}+\Delta^{2}\right]^{1/2}
,\label{uniform1}\\
u({\bf r})&=&(2E)^{-1/2}\left( E+ \hbar^{2}k^{2}/2m-E_{F}\right)^{1/2}
e^{i{\bf k}\cdot{\bf r}} ,\label{uniform2}\\
v({\bf r})&=&(2E)^{-1/2}\left( E-
\hbar^{2}k^{2}/2m+E_{F}\right) ^{1/2} e^{i{\bf
k}\cdot{\bf r}}.\label{uniform3}
\end{eqnarray}
The excitation spectrum is continuous, with excitation gap $\Delta$. The
eigenfunctions $(u,v)$ are plane waves characterized by a wavevector ${\bf k}$.
The coefficients of the plane waves are the two coherence factors of the BCS
(Bardeen-Cooper-Schrieffer) theory.

At an interface between a normal metal and a superconductor the pairing
interaction drops to zero over atomic distances at the normal side. (We assume
non-interacting electrons in the normal region.) Therefore, $\Delta({\bf
r})\equiv 0$ in the normal region. At the superconducting side of the NS
interface, $\Delta({\bf r})$ recovers its bulk value $\Delta$ only at some
distance from the interface. This suppression of $\Delta({\bf r})$ is neglected
in the step-function model
\begin{eqnarray}
\Delta({\bf r})=\left\{\begin{array}{ll}
\Delta&\;{\rm if}\;{\bf r}\in S,\\
0 &\;{\rm if}\;{\bf r}\in N.
\end{array}\right.\label{Delta0}
\end{eqnarray}
The step-function pair potential is also referred to in the literature as a
``rigid boundary condition'' \cite{Lik79}. It greatly simplifies the analysis
of the problem without changing the results in any qualitative way.

Since we will only be considering a single superconductor, the phase of the
superconducting order parameter is irrelevant and we may take $\Delta$ real.
(See Ref.\ \cite{Bee92b} for a tutorial introduction to mesoscopic Josephson
junctions, such as a quantum dot connected to two superconductors.)

\section{Minigap in NS junctions}
\label{minigap}

The presence of a normal metal interface changes the excitation spectrum
(proximity effect). The continuous spectrum above the bulk gap $\Delta$ differs
from the BCS form (\ref{uniform2}) and in addition there may appear discrete
energy levels $E_{n}<\Delta$.

The wave function of the lowest level contains electron and hole components
$u_{0},v_{0}$ of equal magnitude, mixed by Andreev reflection. The mean time
$\tau_{\rm dwell}$ between Andreev reflections (corresponding to the mean life
time of an electron or hole excitation) sets the scale $E_{0}\equiv E_{\rm
gap}\simeq\hbar/\tau_{\rm dwell}$ for the energy of this lowest level
\cite{McM68}. This ``minigap'' is smaller than the bulk gap by a factor
$\xi_{0}/v_{F}\tau_{\rm dwell}$, with $\xi_{0}=\hbar v_{F}/\Delta$ the
superconducting coherence length and $v_{F}$ the Fermi velocity. The energy
$\hbar/\tau_{\rm dwell}$ is called the Thouless energy $E_{T}$, because of the
role it plays in Thouless's theory of localization \cite{Imr02}.

The simplest NS junction, which can be analyzed exactly \cite{DeG63}, consists
of an impurity-free normal metal layer (thickness $d$) on top of a bulk
superconductor. Because of translational invariance parallel to the NS
interface, the parallel component $p_{\parallel}$ of the momentum is a good
quantum number. The lowest excitation energy
\begin{equation}
E_{0}(p_{\parallel})=\frac{\pi\hbar}{2T(p_{\parallel})},\;\;
T(p_{\parallel})=\frac{2md}{(p_{F}^{2}-p_{\parallel}^{2})^{1/2}},\label{E0ppar}
\end{equation}
is the reciprocal of the time $T(p_{\parallel})$ between two subsequent Andreev
reflections. This time diverges when $p_{\parallel}$ approaches the Fermi
momentum $p_{F}=\hbar k_{F}=\sqrt{2mE_{F}}$, so $E_{0}$ can come
microscopically close to zero. The lower limit $E_{0}\agt\hbar^{2}/md^{2}$ is
set by the quantization of the momentum perpendicular to the layer.

\begin{figure}
\centerline{\includegraphics[width=8cm]{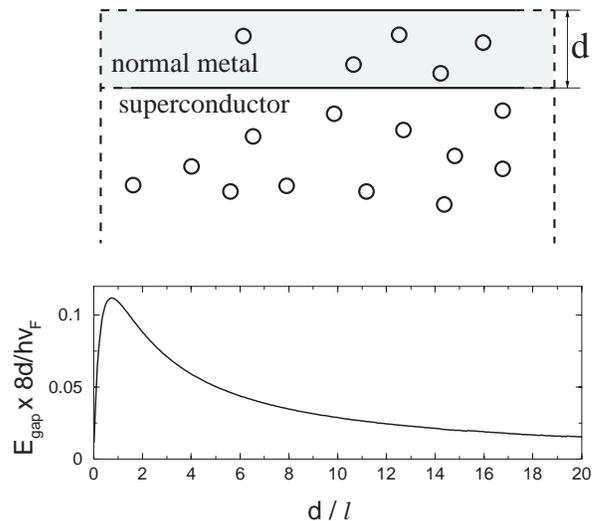}}
\caption{
Excitation gap $E_{\rm gap}$ of a disordered NS junction, as a function of the
ratio of the thickness $d$ of the normal metal layer and the mean free path
$l$. The curve in the bottom panel is calculated from the disorder-averaged
Green function (for $\xi_{0}\ll d,l$). The top panel illustrates the geometry.
The normal metal layer has a specularly reflecting upper surface and an ideally
transmitting lower surface. Adapted from Ref.\ \cite{Pil00}.
\label{pilgram}
}
\end{figure}

Impurities in the normal metal layer (with mean free path $l$) prevent the time
between Andreev reflections to grow much larger than $\tau_{\rm dwell}\simeq
\max(l/v_{F},d^{2}/v_{F}l)$. The excitation gap \cite{Gol88,Bel96,Pil00}
\begin{equation}
E_{\rm gap}\simeq \hbar/\tau_{\rm dwell}\simeq (\hbar
v_{F}l/d^{2})\min(1,d^{2}/l^{2}) \label{E0l}
\end{equation}
is now a factor $k_{F}l\min(1,d^{2}/l^{2})$ larger than in the absence of
impurities. A precise calculation using disorder-averaged Green functions
(reviewed in Ref.\ \cite{Bel99}) gives the curve shown in Fig.\ \ref{pilgram}.
The two asymptotes are \cite{Pil00}
\begin{equation}
E_{\rm gap}=\left\{\begin{array}{cc}
0.43\,\hbar v_{F}/l,&{\rm if}\;\;d/l\ll 1,\\
0.78\,\hbar D/d^{2},&{\rm if}\;\;d/l\gg 1,
\end{array}\right.\label{EgapNS}
\end{equation}
with $D=v_{F}l/3$ the diffusion constant in the normal metal.

The minigap in a ballistic quantum dot (Andreev billiard) differs from that in
a disordered NS junction in two qualitative ways:
\begin{enumerate}
\item The opening of an excitation gap depends on the shape of the boundary,
rather than on the degree of disorder \cite{Mel96}. A chaotic billiard has a
gap at the Thouless energy $E_{T}\simeq\hbar/\tau_{\rm dwell}$, like a
disordered NS junction. An integrable billiard has a linearly vanishing density
of states, like a ballistic NS junction.
\item In a chaotic billiard a new time scale appears, the Ehrenfest time
$\tau_{E}$, which competes with $\tau_{\rm dwell}$ in setting the scale for the
excitation gap \cite{Lod98}. While $\tau_{\rm dwell}$ is a classical
$\hbar$-independent time scale, $\tau_{E}\propto|\ln\hbar|$ has a quantum
mechanical origin.
\end{enumerate}

Because one can not perform a disorder average in Andreev billiards, the Green
function formulation is less useful than in disordered NS junctions. Instead,
we will make extensive use of the scattering matrix formulation, explained in
the next section.

\section{Scattering formulation}
\label{scatteringformu}

In the step-function model (\ref{Delta0}) the excitation spectrum of the
coupled electron-hole quasiparticles can be expressed entirely in terms of the
scattering matrix of normal electrons \cite{Bee91}.

\begin{figure}
\centerline{\includegraphics[width=6cm]{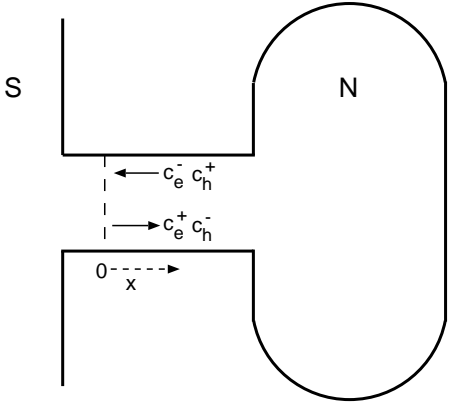}}
\caption{Normal metal (N) containing an Andreev billiard, coupled to a
superconductor (S) by an ideal lead. The dashed line represents the NS
interface. Scattering states $c^{\rm in}=(
c_{\rm e}^{+},c_{\rm h}^{-})$ and $c^{\rm out}=(
c_{\rm e}^{-},c_{\rm h}^{+})$ are indicated schematically.
\label{NSgeometry}
}
\end{figure}

The scattering geometry is illustrated in Fig.\ \ref{NSgeometry}. It consists
of a finite normal-metal region N adjacent to a semi-infinite superconducting
region S. The metal region represents the Andreev billiard. To obtain a
well-defined scattering problem we insert an ideal (impurity-free) normal lead
between N and S. We assume that the only scattering in the superconductor
consists of Andreev reflection at the NS interface (no disorder in S). The
superconductor may then also be represented by an ideal lead. We choose a
coordinate system so that the normal and superconducting leads lie along the
$x$-axis, with the interface at $x=0$.

We first construct a basis for the scattering matrix. In the normal lead N the
eigenfunctions of the BdG equation (\ref{BdG1}) can be written in the form
\begin{subequations}
\label{PsiN}
\begin{eqnarray}
&&\Psi_{n,{\rm e}}^{\pm}({\rm N})=
{\renewcommand{\arraystretch}{0.6}
\left(\begin{array}{c}1\\ 0\end{array}\right)}
\frac{1}{\sqrt{k_{n}^{\rm e}}}\,\Phi_{n}(y,z)
\exp(\pm{\rm i}k_{n}^{\rm e}x),\label{PsiNa}\\
&&\Psi_{n,{\rm h}}^{\pm}({\rm N})=
{\renewcommand{\arraystretch}{0.6}
\left(\begin{array}{c}0\\ 1\end{array}\right)}
\frac{1}{\sqrt{k_{n}^{\rm h}}}\,\Phi_{n}(y,z)
\exp(\pm{\rm i}k_{n}^{\rm h}x),\label{PsiNb}
\end{eqnarray}
\end{subequations}
where the wavenumbers $k_{n}^{\rm e}$ and $k_{n}^{\rm h}$ are given by
\begin{eqnarray}
k_{n}^{\rm e,h}= \frac{\sqrt{2m}}{\hbar}(E_{F}
-E_{n}+\sigma^{\rm e,h}E)^{1/2}, \label{keh}
\end{eqnarray}
and we have defined $\sigma^{\rm e}\equiv 1$,
$\sigma^{\rm h}\equiv -1$. The labels e and h indicate
the electron or hole character of the wave function.
The index $n$ labels the modes, $\Phi_{n}(y,z)$ is the transverse wave function
of the $n$-th mode, and $E_{n}$ its threshold energy:
\begin{eqnarray}
[(p_{y}^{2}+p_{z}^{2})/2m+V(y,z)]\Phi_{n}(y,z)=E_{n}\Phi_{n}(y,z).\label{Phin}
\end{eqnarray}
The eigenfunction $\Phi_{n}$ is normalized to unity, $\int\! dy\int\! dz
\,|\Phi_{n}|^{2}=1$.

In the superconducting lead S the eigenfunctions are
\begin{subequations}
\label{PsiS}
\begin{eqnarray}
\Psi_{n,{\rm e}}^{\pm}({\rm S})&=&
{\renewcommand{\arraystretch}{0.6}
\left(\begin{array}{c}
{\rm e}^{{\rm i}\eta^{\rm e}/2}\\
{\rm e}^{-{\rm i}\eta^{\rm e}/2}\end{array}\right)}
\frac{1}{\sqrt{2q_{n}^{\rm e}}}
(E^{2}/\Delta^{2}-1)^{-1/4}\nonumber\\
&&\mbox{}\times\Phi_{n}(y,z)\exp(\pm{\rm i}q_{n}^{\rm e}x),\label{PsiSa}\\
\Psi_{n,{\rm h}}^{\pm}({\rm S})&=&
{\renewcommand{\arraystretch}{0.6}
\left(\begin{array}{c}
{\rm e}^{{\rm i}\eta^{\rm h}/2}\\
{\rm e}^{-{\rm i}\eta^{\rm h}/2}\end{array}\right)}
\frac{1}{\sqrt{2q_{n}^{\rm h}}}
(E^{2}/\Delta^{2}-1)^{-1/4}\nonumber\\
&&\mbox{}\times\Phi_{n}(y,z)\exp(\pm{\rm i}q_{n}^{\rm h}x).
\label{PsiSb}
\end{eqnarray}
\end{subequations}
We have defined
\begin{eqnarray}
q_{n}^{\rm e,h}&= &\frac{\sqrt{2m}}{\hbar}
[E_{F}-E_{n}+\sigma^{\rm e,h}
(E^{2}-\Delta^{2}) ^{1/2}]^{1/2},\label{qneh}\\
\eta^{\rm e,h}&= &\sigma^{\rm e,h}
\arccos(E/\Delta).\label{etaeh}
\end{eqnarray}

The wave functions (\ref{PsiN}) and (\ref{PsiS}) have
been normalized to carry the same amount of quasiparticle
current, because we want to use them as the basis for a unitary
scattering matrix. The direction of the velocity is the same as the wave vector
for the electron and opposite for the hole.

A wave incident on the Andreev billiard is described in the basis (\ref{PsiN})
by a vector of coefficients
\begin{eqnarray}
c^{\rm in}=(
c_{\rm e}^{+},c_{\rm h}^{-}),\label{cNin}
\end{eqnarray}
as shown schematically in Fig.\ \ref{NSgeometry}. (The mode index $n$ has been
suppressed for simplicity of notation.) The reflected wave has vector of
coefficients
\begin{eqnarray}
c^{\rm out}=(
c_{\rm e}^{-},c_{\rm h}^{+}).\label{cNout}
\end{eqnarray}
The scattering matrix $S_{\rm N}$ of the normal region
relates these two vectors, $c_{\rm N}^{\rm out}=
S_{\rm N}c_{\rm N}^{\rm in}$. Because the normal
region does not couple electrons and holes, this
matrix has the block-diagonal form
\begin{eqnarray}
S_{\rm N}(E)=
{\renewcommand{\arraystretch}{0.6}
\left(\begin{array}{cc}
S(E)&0\\
\!0&S(-E)^{\ast}
\end{array}\right)}.
\label{sN}
\end{eqnarray}
Here $S(E)$ is the unitary scattering matrix associated
with the single-electron Hamiltonian $H$. It is an $N\times N$ matrix, with
$N(E)$ the number of propagating modes at energy $E$. The dimension of $S_{\rm
N}(E)$ is $N(E)+N(-E)$.

For energies $0<E<\Delta$ there are no propagating modes in the
superconducting lead S. Restricting ourselves to that energy range, we
can define a scattering matrix $S_{\rm A}$ for Andreev reflection at the
NS interface by $c^{\rm in} =S_{\rm A}c^{\rm out}$. The
elements of $S_{\rm A}$ are obtained by matching the wave function
(\ref{PsiN}) at $x=0$ to the decaying wave function (\ref{PsiS}). Since
$\Delta\ll E_{F}$ one may ignore normal reflections at the NS
interface and neglect the difference between $N(E)$ and $N(-E)$. This
is known as the Andreev approximation \cite{And64}. The result is
\begin{eqnarray}
&&S_{\rm A}(E)=
{\renewcommand{\arraystretch}{0.6}
\left(\begin{array}{cc}
0&\alpha(E)\\
\alpha(E)&0
\end{array}\right)},
\label{sA}\\
&&\alpha(E)=e^{-i\arccos(E/\Delta)}=\frac{E}{\Delta}-
i\sqrt{1-\frac{E^{2}}{\Delta^{2}}}. \label{alphadef}
\end{eqnarray}
Andreev reflection transforms an electron mode into a hole mode,
without change of mode index. The transformation is accompanied by a
phase shift $-\arccos(E/\Delta)$ due to the penetration
of the wave function into the superconductor.

We are now ready to relate the excitation spectrum
of the Andreev billiard to the scattering matrix of the normal region. We
restrict ourselves to the discrete spectrum (see Ref.\ \cite{Bee91} for the
continuous spectrum). The condition $c_{\rm in}=
S_{\rm A}S_{\rm N}c_{\rm in}$ for a bound state implies
${\rm Det}\,(1-S_{\rm A}S_{\rm N})=0$. Using
Eqs.\ (\ref{sN}), (\ref{sA}), and the identity
\begin{eqnarray}
{\rm Det}\,
{\renewcommand{\arraystretch}{0.4}
\left(\begin{array}{cc}
a&b\\c&d
\end{array}\right)}={\rm Det}\,(ad-aca^{-1}b)
\label{identity}
\end{eqnarray}
one obtains the equation \cite{Bee91}
\begin{equation}
{\rm Det}\,\left[1-\alpha(E)^{2}S(E)S(-E)^{\ast}\right]=0.\label{discreteE}
\end{equation}
The roots $E_{p}$ of this determinantal equation constitute the discrete
spectrum of the Andreev billiard.

\section{Stroboscopic model}
\label{strobomodel}

Although the phase space of the Andreev billiard is four-dimensional, like for
any billiard it can be reduced to two dimensions on a Poincar\'{e} surface of
section \cite{Gut90,Haa01}. This amounts to a stroboscopic description of the
classical dynamics, because the position and momentum are only recorded when
the particle crosses the surface of section. Quantum mechanically, the
stroboscopic evolution of the wave function is described by a compact unitary
map rather than by a noncompact Hermitian operator \cite{Bog92,Pra03}. What one
loses by the stroboscopic description is information on time scales below the
time of flight across the billiard. What one gains is an enormous increase in
computational efficiency.

A stroboscopic model of an Andreev billiard was constructed by Jacquod et al.\
\cite{Jac03}, building on an existing model for open normal billiards called
the open kicked rotator \cite{Oss02}. The Andreev kicked rotator possesses the
same phenomenology as the Andreev billiard, but is much more tractable
numerically.\footnote{The largest simulation to date of a two-dimensional
Andreev billiard has $N=30$, while for the Andreev kicked rotator $N=10^{5}$ is
within reach, cf.\ Fig.\ \protect\ref{comparison2}.}
In this subsection we discuss how it is formulated. Some results obtained by
this numerical method will be compared in subsequent sections with results
obtained by analytical means.

A compact unitary map is represented in quantum mechanics by the
Floquet operator $F$, which gives the stroboscopic time evolution
$u(p\tau_0)=F^{p}u(0)$ of an initial wave function $u(0)$. (We set
the stroboscopic period $\tau_0=1$ in most equations.) The unitary
$M\times M$ matrix $F$ has eigenvalues $\exp(-i\varepsilon_{m})$, with
the quasi-energies $\varepsilon_{m}\in (-\pi,\pi)$ (measured in units
of $\hbar/\tau_{0}$). This describes the electron excitations above the
Fermi level. Hole excitations below the Fermi level have Floquet operator
$F^{\ast}$ and wave function $v(p)=(F^{\ast})^{p}v(0)$. The mean level
spacing of electrons and holes separately is $\delta=2\pi/M$.

An electron is converted into a hole by Andreev
reflection at the NS interface, with phase shift $-i$ for $\varepsilon\ll
\tau_{0}\Delta/\hbar$ [cf.\ Eq.\ (\ref{alphadef})]. In the stroboscopic
description one assumes that Andreev reflection occurs only at times which are
multiples of $\tau_{0}$. The $N\times M$ matrix $P$ projects onto the NS
interface. Its elements are $P_{nm}=1$ if $m=n \in \{n_{1},n_{2},\ldots
n_{N}\}$ and $P_{nm}=0$ otherwise.  The dwell time of a quasiparticle
excitation in the normal metal is $\tau_{\rm dwell}=M/N$, equal to the
mean time between Andreev reflections.

Putting all this together one constructs the quantum Andreev map from the
matrix product
\begin{equation}
{\cal F}={\cal P}
\left(\begin{array}{cc}
F&0\\
0&F^{\ast}
\end{array}\right),\;\;
{\cal P}=\left(\begin{array}{cc}
1-P^{\rm T}P&-iP^{\rm T}P\\
-iP^{\rm T}P&1-P^{\rm T}P
\end{array}\right).\label{calFdef}
\end{equation}
(The superscript ``T'' indicates the transpose of a matrix.) The
particle-hole wave function $\Psi=(u,v)$ evolves in time as $\Psi(p)={\cal
F}^{p}\Psi(0)$. The Floquet operator can be symmetrized (without changing
its eigenvalues) by the unitary transformation ${\cal F}\rightarrow
{\cal P}^{-1/2}{\cal F}{\cal P}^{1/2}$, with
\begin{equation}
{\renewcommand{\arraystretch}{1.3}
{\cal P}^{1/2}=\left(\begin{array}{cc}
1-(1-{\textstyle\frac{1}{2}}\sqrt{2})P^{\rm
T}P&-i{\textstyle\frac{1}{2}}\sqrt{2}P^{\rm T}P\\
-i{\textstyle\frac{1}{2}}\sqrt{2}P^{\rm
T}P&1-(1-{\textstyle\frac{1}{2}}\sqrt{2})P^{\rm T}P
\end{array}\right).\label{calFsymdef}
}
\end{equation}

The quantization condition ${\rm det}({\cal F}-e^{-i\varepsilon})=0$
can be written equivalently as \cite{Jac03}
\begin{equation}
{\rm Det}\,[1+S(\varepsilon)S(-\varepsilon)^{\ast}]=0,\label{detS}
\end{equation}
in terms of the $N\times N$ scattering matrix \cite{Oss02,Fyo00}
\begin{equation}
S(\varepsilon)=P[e^{-i\varepsilon}-F(1-P^{\rm T}P)]^{-1}
FP^{\rm T}.\label{Sdef}
\end{equation}
Eq.\ (\ref{detS}) for the Andreev map has the same form as Eq.\
(\ref{discreteE}) for the
Andreev billiard (with $\alpha\rightarrow -i$). In particular, both equations
have roots that lie symmetrically around zero.

A specific realization of the Andreev map is the Andreev kicked rotator. (See
Ref.\ \cite{Oss04} for a different realization, based on the kicked Harper
model.) The normal kicked rotator has Floquet operator \cite{Izr90}
\begin{eqnarray}
F&=&\exp\left(i\frac{\hbar\tau_{0}}{4I_{0}}
\frac{\partial^{2}}{\partial\theta^{2}} \right)\exp\left(-i\frac{KI_{0}}{\hbar\tau_{0}}\cos\theta\right)
\nonumber\\
&&\mbox{}\times\exp\left(i\frac{\hbar\tau_{0}}{4I_{0}}
\frac{\partial^{2}}{\partial\theta^{2}} \right). \label{kickedF}
\end{eqnarray}
It describes a particle that moves freely along the unit circle
$(\cos\theta,\sin\theta)$ with moment of inertia $I_{0}$ for half
a period $\tau_{0}$, is then kicked with a strength $K\cos\theta$,
and proceeds freely for another half period. Upon increasing $K$ the
classical dynamics varies from fully integrable ($K=0$) to fully chaotic
[$K\agt 7$, with Lyapunov exponent $\alpha\approx\ln (K/2)$]. For $K<7$
stable and unstable motion coexist (mixed phase space). If needed, a magnetic
field can be introduced into the model as described in Ref.\ \cite{Two04}.

The transition from classical to quantum behavior is governed by the
effective Planck constant $h_{\rm eff}\equiv\hbar\tau_{0}/2\pi I_{0}$.
For $1/h_{\rm eff}\equiv M$ an even integer, $F$ can be represented
by an $M\times M$ unitary symmetric matrix. The angular coordinate
and momentum eigenvalues are $\theta_{m}=2\pi m/M$ and $p_{m}=\hbar
m$, with $m=1,2,\ldots M$, so phase space has the topology of a
torus. The NS interface is an annulus around the torus, either in the
$\theta$-direction or in the $p$-direction. (The two configurations give
equivalent results.) The construction (\ref{calFdef}) produces a $2M\times
2M$ Floquet operator ${\cal F}$, which can be diagonalized efficiently
in ${\cal O}(M^2 \ln M)$ operations [rather than ${\cal O}(M^3)$]
by combining the Lanczos technique
with the fast-Fourier-transform algorithm \cite{Ket99}.

\section{Random-matrix theory}
\label{RMT}

An ensemble of isolated chaotic billiards, constructed by varying the shape at
constant area, corresponds to an ensemble of Hamiltonians $H$ with a particular
distribution function $P(H)$. It is convenient to think of the Hamiltonian as a
random $M\times M$ Hermitian matrix, eventually sending $M$ to infinity. The
basic postulate of random-matrix theory (RMT) \cite{Meh91} is that the
distribution is invariant under the unitary transformation $H\rightarrow
UHU^{\dagger}$, with $U$ an arbitrary unitary matrix. This implies a
distribution of the form
\begin{equation}
P(H)\propto\exp[-{\rm Tr}\,V(H)].\label{WDPH}
\end{equation}
If $V(H)\propto H^{2}$, the ensemble is called Gaussian. This choice simplifies
some of the calculations but is not essential, because the spectral
correlations become largely independent of $V$ in the limit
$M\rightarrow\infty$. More generally, the ensemble of the form (\ref{WDPH}) is
called the Wigner-Dyson ensemble, after the founding fathers of RMT.

By computing the Jacobian from the space of matrix elements to the space of
eigenvalues $E_{n}$ ($n=1,2,\ldots M$), one obtains the eigenvalue probability
distribution \cite{Meh91}
\begin{equation}
P(\{E_n\})\propto \prod_{i<j} |E_i-E_j|^\beta
\prod_k e^{-V(E_{k})}.\label{eigenvaluedist}
\end{equation}
The symmetry index $\beta$ counts the number of degrees of freedom in the
matrix elements. These are real ($\beta=1$) in the presence of time-reversal
symmetry or complex ($\beta=2$) in its absence. (A third possibility,
$\beta=4$, applies to time-reversally symmetric systems with strong spin-orbit
scattering, which we will not consider here.) Since the unitary transformation
$H\rightarrow UHU^{\dagger}$ requires an orthogonal $U$ to keep a real
Hamiltonian, one speaks of the Gaussian orthogonal ensemble (GOE) when
$\beta=1$. The name Gaussian unitary ensemble (GUE) refers to $\beta=2$.

There is overwhelming numerical evidence that chaotic billiards are well
described by the Wigner-Dyson ensemble \cite{Haa01}. (This is known as the
Bohigas-Giannoni-Schmit conjecture \cite{Boh84}.) A complete theoretical
justification is still lacking, but much progress has been made in that
direction \cite{Mul04}. In this section we will take Eq.\ (\ref{WDPH}) for the
ensemble of isolated billiards as our starting point and deduce what properties
it implies for the ensemble of Andreev billiards.

\begin{figure}
\centerline{\includegraphics[width=6cm]{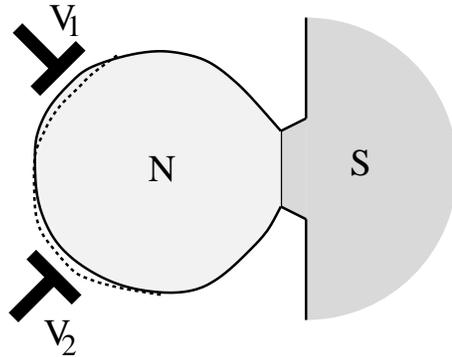}}
\caption{
A quantum dot (N) connected to a superconductor (S). The voltages on the gates
$V_{1}$ and $V_{2}$ change the shape of the dot. Different values of the
applied voltages create different samples within the same ensemble. From Ref.\
\cite{Vav01}.
\label{dotwithgates}
}
\end{figure}

The isolated billiard becomes an Andreev billiard when it is connected by a
point contact to a superconductor, cf.\ Fig.\ \ref{dotwithgates}. In the
isolated billiard RMT breaks down on energy scales greater than
$\hbar/\tau_{\rm erg}$, with the ergodic time $\tau_{\rm erg}\simeq
A^{1/2}/v_{F}$ set by the time of flight across the billiard (of area $A$, at
Fermi velocity $v_{F}$). On larger energy scales, hence on shorter time scales,
non-chaotic dynamics appears which is beyond RMT. The superconductor affects
the billiard in an energy range around the Fermi level that is set by the
Thouless energy $E_{T}\simeq\hbar/\tau_{\rm dwell}$. (We assume that $E_{T}$ is
less than the gap $\Delta$ in the bulk superconductor.) In this context the
dwell time $\tau_{\rm dwell}$ is the mean time between Andreev reflections
(being the life time of an electron or hole quasiparticle). The condition
$\tau_{\rm erg}\ll\tau_{\rm dwell}$ of weak coupling is therefore sufficient to
be able to apply RMT to the entire relevant energy range.

\subsection{Effective Hamiltonian}
\label{RMTeffH}

The excitation energies $E_{p}$ of the Andreev billiard in the discrete part of
the spectrum are the solutions of the determinantal equation (\ref{discreteE}),
given in terms of the scattering matrix $S(E)$ in the normal state (i.e.\ when
the superconductor is replaced by a normal metal). This equation can
alternatively be written in terms of the Hamiltonian $H$ of the isolated
billiard and the $M\times N$ coupling matrix $W$ that describes the $N$-mode
point contact. The relation between $S$ and $H,W$ is \cite{Guh98,Bee97}
\begin{equation}
S(E)=1-2\pi i W^{T}(E-H+i\pi WW^{T})^{-1}W.\label{SWHformula}
\end{equation}
The $N\times N$ matrix $W^{T}W$ has eigenvalues $w_{n}$ given by
\begin{equation}
w_{n}=\frac{M\delta}{\pi^{2}\Gamma_{n}}
\left(2-\Gamma_{n}-2\sqrt{1-\Gamma_{n}}\right), \label{Wgammarelation}
\end{equation}
where $\delta$ is the mean level spacing in the isolated billiard and
$\Gamma_{n}\in[0,1]$ is the transmission probability of mode $n=1,2,\ldots N$
in the point contact. For a ballistic contact, $\Gamma_{n} = 1$, while
$\Gamma_{n} \ll 1$ for a tunneling contact. Both the number of modes $N$ and
the level spacing $\delta$ refer to a single spin direction.

Substituting Eq.\ (\ref{SWHformula}) into Eq.\ (\ref{discreteE}), one arrives
at an alternative determinantal equation for the discrete spectrum
\cite{Bro97a}:
\begin{eqnarray}
&&{\rm Det}\,[E-{\cal H}+{\cal W}(E)]=0,\label{DetEHW}\\
&&{\cal H}=\left(\begin{array}{cc}
H&0\\0&-H^{\ast}
\end{array}\right),\label{calHdef}\\
&&{\cal W}(E)=
\frac{\pi}{\sqrt{\Delta^{2}-E^{2}}}
\left(\begin{array}{cc}
EWW^{T}&\Delta WW^{T}\\
\Delta WW^{T}&EWW^{T}
\end{array}\right).\label{calWdef}
\end{eqnarray}
The density of states follows from
\begin{equation}
\rho(E)=-\frac{1}{\pi}{\rm Im}\,{\rm Tr}\,(1+d{\cal W}/dE)(E + i0^{+}-{\cal
H}+{\cal W})^{-1}.\label{resolvent}
\end{equation}

In the relevant energy range $E\alt E_{T}\ll\Delta$ the matrix ${\cal W}(E)$
becomes energy independent. The excitation energies can then be obtained as the
eigenvalues of the effective Hamiltonian \cite{Fra96}
\begin{equation}
{\cal H}_{\rm eff}=\left(\begin{array}{cc}
H&-\pi WW^{T}\\-\pi WW^{T}&-H^{\ast}
\end{array}\right).\label{calHeffdef}
\end{equation}
The effective Hamiltonian ${\cal H}_{\rm eff}$ should not be confused with the
Bogoliubov-de Gennes Hamiltonian ${\cal H}_{\rm BG}$, which contains the
superconducting order parameter in the off-diagonal blocks [cf.\ Eq.\
(\ref{HBG})]. The Hamiltonian ${\cal H}_{\rm BG}$ determines the entire
excitation spectrum (both the discrete part below $\Delta$ and the continuous
part above $\Delta$), while the effective Hamiltonian ${\cal H}_{\rm eff}$
determines only the low-lying excitations $E_{p}\ll\Delta$.

The Hermitian matrix ${\cal H}_{\rm eff}$ (like ${\cal H}_{\rm BG}$) is
antisymmetric under the combined operation of charge conjugation (${\cal C}$)
and time inversion (${\cal T}$) \cite{Alt96}:
\begin{equation}
{\cal H}_{\rm eff}=-\sigma_{y}{\cal H}_{\rm eff}^{T}\sigma_{y},\;\;
\sigma_{y}=
\left(\begin{array}{cc}
0&-i\\
i& 0\\
\end{array}\right).\label{CTsymmetry}
\end{equation}
(An $M\times M$ unit matrix in each of the four blocks of $\sigma_{y}$ is
implicit.) The ${\cal CT}$-antisymmetry ensures that the eigenvalues lie
symmetrically around $E=0$. Only the positive eigenvalues are retained in the
excitation spectrum, but the presence of the negative eigenvalues is felt as a
level repulsion near $E=0$.

\subsection{Excitation gap}
\label{RMTexcgap}

In zero magnetic field the suppression of the density of states $\rho(E)$
around $E=0$ extends over an energy range $E_{T}$ that may contain many level
spacings $\delta$ of the isolated billiard. The ratio $g\simeq E_{T}/\delta$ is
the conductance of the point contact in units of the conductance quantum
$e^{2}/h$. For $g\gg 1$ the excitation gap $E_{\rm gap}\simeq g\delta$ is a
mesoscopic quantity, because it is intermediate between the microscopic energy
scale $\delta$ and the macroscopic energy scale $\Delta$. One can use
perturbation theory in the small parameter $1/g$ to calculate $\rho(E)$. The
analysis presented here follows the RMT of Melsen et al. \cite{Mel96}. An
alternative derivation \cite{Vav03}, using the disorder-averaged Green
function, is discussed in the next sub-section.

In the presence of time-reversal symmetry the Hamiltonian $H$ of the isolated
billiard is a real symmetric matrix. The appropriate RMT ensemble is the GOE,
with distribution \cite{Meh91}
\begin{equation}
P(H) \propto\exp\left(-\frac{\pi^{2}}{4M\delta^{2}}{\rm
Tr}\,H^{2}\right).\label{PGOE}
\end{equation}
The ensemble average $\langle\cdots\rangle$ is an average over $H$ in the GOE
at fixed coupling matrix $W$. Because of the block structure of ${\cal H}_{\rm
eff}$, the ensemble averaged Green function ${\cal G}(E)=\langle(E-{\cal
H}_{\rm eff})^{-1}\rangle$ consists of four $M\times M$ blocks ${\cal G}_{11}$,
${\cal G}_{12}$, ${\cal G}_{21}$, ${\cal G}_{22}$. By taking the trace of each
block separately, one arrives at a $2\times 2$ matrix Green function
\begin{equation}
G=\left(\begin{array}{cc} G_{11} & G_{12}\\G_{21} & G_{22}\end{array}
 \right) =\frac{\delta}{\pi}\left(\begin{array}{cc}
{\rm Tr}\,{\cal G}_{11}&{\rm Tr}\,{\cal G}_{12}\\
{\rm Tr}\,{\cal G}_{21}&{\rm Tr}\,{\cal G}_{22}
\end{array}\right).\label{Gmatrixdef}
\end{equation}
(The factor $\delta/\pi$ is inserted for later convenience.)

The average over the distribution (\ref{PGOE}) can be done diagrammatically
\cite{Pan81,Brez94}. To leading order in $1/M$ and for $E\gg\delta$ only simple
(planar) diagrams need to be considered. Resummation of these diagrams leads to
the selfconsistency equation \cite{Mel96,Bro97a}
\begin{equation}
{\cal G}=[E+{\cal W}-(M\delta/\pi)\sigma_{z}G\sigma_{z}]^{-1},\;\;
\sigma_{z}=
\left(\begin{array}{cc}
1&0\\
0&-1\\
\end{array}\right).\label{pastur1}
\end{equation}
This is a matrix-generalization of Pastur's equation in the RMT of normal
systems \cite{Pas72}.

The matrices in Eq.\ (\ref{pastur1}) have four $M\times M$ blocks. By taking
the trace of each block one obtains an equation for a $4\times 4$ matrix,
\begin{eqnarray}
&&G=\frac{1}{M}\sum_{m=1}^{M} \left(\begin{array}{cc}
\pi E/M\delta-G_{11}&\tilde{w}_{m}+G_{12}\\
\tilde{w}_{m}+G_{21}&\pi E/M\delta-G_{22}
\end{array}\right)^{-1},\label{pastur}\\
&&\tilde{w}_{m}=\left\{\begin{array}{cc}
\pi^{2}w_{m}/M\delta&{\rm if}\;\;m=1,2,\ldots N,\\
0&{\rm if}\;\;m=N+1,\ldots M.
\end{array}\right.\label{wtildedef}
\end{eqnarray}
Since $G_{22}=G_{11}$ and $G_{21}=G_{12}$ there are two unknown functions to
determine. For $M\gg N$ these satisfy
\begin{subequations}
\label{dysongamma}
\begin{eqnarray}
&&G_{12}^{2}=1+G_{11}^{2},\label{dysongamma1}\\
&&\frac{2\pi
E}{\delta}G_{12}=G_{11}\sum_{n=1}^{N}(-G_{12}+1-2/\Gamma_{n})^{-1},
\label{dysongamma2}
\end{eqnarray}
\end{subequations}
where we have used the relation (\ref{Wgammarelation}) between the parameters
$w_{n}$ and the transmission probabilities $\Gamma_{n}$. Eq.\
(\ref{dysongamma}) has multiple solutions. The physical solution satisfies
$\lim_{E\rightarrow\infty}\langle\rho(E)\rangle=2/\delta$, when substituted
into
\begin{equation}
\langle\rho(E)\rangle=-(2/\delta)\,{\rm Im}\,G_{11}(E).\label{onemoretrace}
\end{equation}

\begin{figure}
\centerline{\includegraphics[width=8cm]{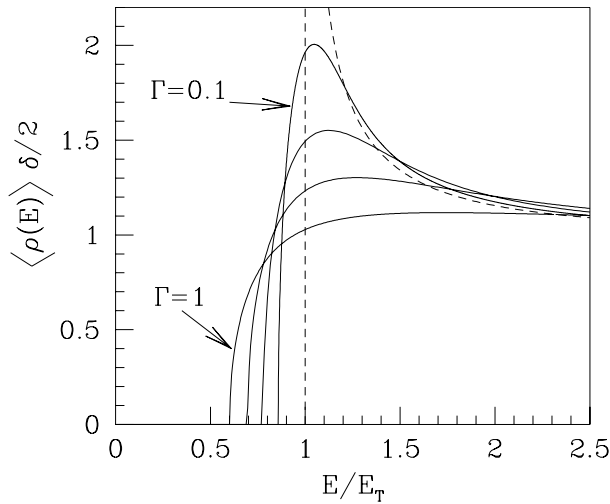}}
\caption{Ensemble averaged density of states of a chaotic billiard coupled by a
point contact to a superconductor, for several values of the transmission
probability through the point contact. The energy is in units of the Thouless
energy $E_{T}=N\Gamma\delta/4\pi$. The solid curves are computed from Eqs.\
(\ref{dysongamma}) and (\ref{onemoretrace}), for mode-independent transmission
probabilities $\Gamma = 1$, $0.5$, $0.25$, $0.1$. The dashed curve is the
asymptotic result (\ref{rhoEtunnel}) for $\Gamma \ll 1$. Adapted from Ref.\
\cite{Mel96}. (The definition of $\delta$ used in that paper differs from the
one used here by a factor of two.)
\label{RMTDOS}
}
\end{figure}

In Fig.\ \ref{RMTDOS} we plot the density of states in the mode-independent
case $\Gamma_{n}\equiv\Gamma$, for several vaues of $\Gamma$. It vanishes as a
square root near the excitation gap. The value of $E_{\rm gap}$ can be
determined directly by solving Eq.\ (\ref{dysongamma}) jointly with
$dE/dG_{11}=0$. The result is
\begin{eqnarray}
&&\frac{k^6-k^4}{(1-k)^6}x^6-\frac{3k^4-20k^2+16}{(1-k)^4}x^4+
\frac{3k^2+8}{(1-k)^2} x^2=1,\nonumber\\
&&x=E_{\rm gap}/E_{T},\;
k=1-2/\Gamma,\;E_{T}=N\Gamma\delta/4\pi.\label{gapequation}
\end{eqnarray}
For later use we parametrize the square-root dependence near the gap as
\begin{equation}
\langle\rho(E)\rangle\rightarrow\frac{1}{\pi}\sqrt{\frac{E-E_{\rm
gap}}{\Delta^3_{\rm gap}}},\;\;E\rightarrow E_{\rm gap}.\label{rhoEneargap}
\end{equation}
When $E\gg E_{\rm gap}$ the density of states approaches the value $2/\delta$
from above, twice the value in the isolated billiard. The doubling of the
density of states occurs because electron and hole excitations are combined in
the excitation spectrum of the Andreev billiard, while in an isolated billiard
electron and hole excitations are considered separately.

A rather simple closed-form expression for $\langle\rho(E)\rangle$ exists in
two limiting cases \cite{Mel96}. In the case $\Gamma=1$ of a ballistic point
contact one has
\begin{eqnarray}
&&\langle\rho(E)\rangle=\frac{E_{T}\sqrt{3}}{3 E \delta} [
Q_{+}(E/E_{T})-Q_{-}(E/E_{T})],\label{rhoEballistica}\\
&&Q_{\pm}(x) = \left[8 - 36 x^2 \pm 3 x \sqrt{3 x^4 + 132 x^2
-48}\right]^{1/3},\label{rhoEballisticb}\\
&&E>E_{\rm gap} = 2\gamma^{5/2} E_{T}=0.60\, E_{T}=\frac{0.30\,\hbar}{\tau_{\rm
dwell}}=0.048\,N\delta,\nonumber\\
&&\label{rhoEballisticc}
\end{eqnarray}
where $\gamma = \frac{1}{2}(\sqrt{5} - 1)$ is the golden number. In this case
the parameter $\Delta_{\rm gap}$ in Eq.\ (\ref{rhoEneargap}) is given by
\begin{equation}
\Delta_{\rm gap}=[(5-2\sqrt{5})\delta^{2}E_{\rm
gap}/8\pi^{2}]^{1/3}=0.068\,N^{1/3}\delta.\label{Deltagap}
\end{equation}
In the opposite tunneling limit $\Gamma \ll 1$ one finds
\begin{equation}
\langle\rho(E)\rangle=\frac{2E}{\delta}({E^2 - E_{T}^2})^{-1/2}, \; E >E_{\rm
gap}= E_{T}.\label{rhoEtunnel}
\end{equation}
In this limit the density of states of the Andreev billiard has the same form
as in the BCS theory for a bulk superconductor \cite{Tin95}, with a reduced
value of the gap (``minigap''). The inverse square-root singularity at the gap
is cut off for any finite $\Gamma$, cf.\ Fig.\ \ref{RMTDOS}.

\subsection{Effect of impurity scattering}
\label{impurity}

Impurity scattering in a chaotic Andreev billiard reduces the magnitude of the
excitation gap by increasing the mean time $\tau_{\rm dwell}$ between Andreev
reflections. This effect was calculated by Vavilov and Larkin \cite{Vav03}
using the method of impurity-averaged Green functions \cite{Bel99}. The minigap
in a disordered quantum dot is qualitatively similar to that in a disordered NS
junction, cf.\ Sec.\ \ref{minigap}. The main parameter is the ratio of the mean
free path $l$ and the width of the contact $W$. (We assume that there is no
barrier in the point contact, otherwise the tunnel probability $\Gamma$ would
enter as well.)

For $l\gg W$ the mean dwell time saturates at the ballistic value
\begin{equation}
\tau_{\rm dwell}=\frac{2\pi\hbar}{N\delta}=\frac{\pi A}{v_{F}W},\;\;{\rm
if}\;\; l\gg W. \label{taudwellbdef}
\end{equation}
In the opposite limite $l\ll W$ the mean dwell time is determined by the
two-dimensional diffusion equation. Up to a geometry-dependent coefficient $c$
of order unity, one has
\begin{equation}
\tau_{\rm dwell}=c\frac{A}{v_{F}l}\ln(A/W^{2}),\;\;{\rm if}\;\; l\ll W.
\label{taudwellddef}
\end{equation}

\begin{figure}
\centerline{\includegraphics[width=8cm]{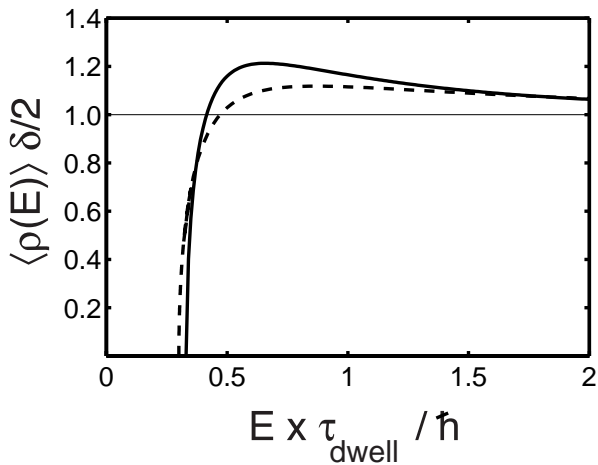}}
\caption{The dashed curve is the $\Gamma=1$ result of Fig.\
\protect\ref{RMTDOS}, corresponding to a quantum dot with weak impurity
scattering (mean free path $l$ much larger than the width $W$ of the point
contact). The solid curve is the corresponding result for strong impurity
scattering ($l\ll W$). The line shape is almost the same, but the energy scale
is different (given by Eqs.\ (\ref{taudwellbdef}) and (\ref{taudwellddef}),
respectively). Adapted from Ref.\ \cite{Vav03}.
\label{usadel}
}
\end{figure}

The density of states in the two limits is shown in Fig.\ \ref{usadel}. There
is little difference, once the energy is scaled by $\tau_{\rm dwell}$. For
$l\gg W$ the excitation gap is given by the RMT result $E_{\rm
gap}=0.300\,\hbar/\tau_{\rm dwell}$, cf.\ Eq.\ (\ref{rhoEballisticc}). For
$l\ll W$ Vavilov and Larkin find $E_{\rm gap}=0.331\,\hbar/\tau_{\rm dwell}$.

\subsection{Magnetic field dependence}
\label{RMTmagnfield}

A magnetic field $B$, perpendicular to the billiard, breaks time-reversal
symmetry, thereby suppressing the excitation gap. A perturbative treatment
remains possible as long as $E_{\rm gap}(B)$ remains large compared to $\delta$
\cite{Mel97}.

The appropriate RMT ensemble for the isolated billiard is described by the
Pandey-Mehta distribution \cite{Meh91,Pan83}
\begin{eqnarray}
P(H)&\propto&\exp\biggl(-\frac{\pi^{2}(1+b^2)}{4M \delta^2}\nonumber\\
&&\mbox{}\times\sum_{i,j=1}^M\left[({\rm Re}\, H_{ij})^2+b^{-2}
({\rm Im}\,H_{ij})^2\right]\biggr).
\label{PandeyMehta}
\end{eqnarray}
The parameter $b \in [0,1]$ measures the strength of the time-reversal symmetry
breaking. The invariance of $P(H)$ under unitary transformations is broken if
$b\neq 0,1$. The relation between $b$ and the magnetic flux $\Phi$ through the
billiard is \cite{Bee97}
\begin{equation}
Mb^2=c (\Phi e/h)^2\frac{\hbar v_{F}}{\delta\sqrt{A}},\label{crossoverpar}
\end{equation}
with $c$ a numerical coefficient that depends only on the shape of the
billiard. Time-reversal symmetry is effectively broken when $Mb^2 \simeq g$,
which occurs for $\Phi\simeq (h/e)\sqrt{\tau_{\rm erg}/\tau_{\rm dwell}}\ll
h/e$. The effect of such weak magnetic fields on the bulk superconductor can be
ignored.

\begin{figure}
\centerline{\includegraphics[width=8cm]{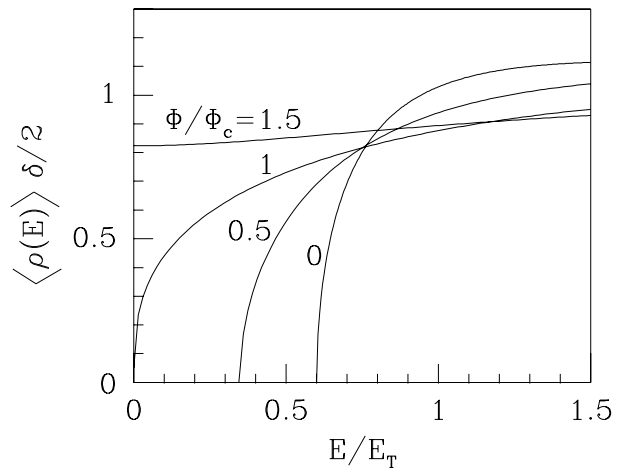}}
\caption{Magnetic field dependence of the density of states for the case of a
ballistic point contact ($\Gamma_{n}\equiv 1$), computed from Eqs.\
(\ref{dysongamma1}), (\ref{onemoretrace}), and (\ref{dysongamma2b}). The
microscopic gap of order $\delta$ which persists when $\Phi>\Phi_{c}$ is not
resolved in this calculation. Adapted from Ref.\ \cite{Mel97}.
\label{RMTDOSB}
}
\end{figure}

The selfconsistency equation for the Green function is the same as Eq.\
(\ref{pastur}), with one difference: On the right-hand-side the terms $G_{12}$
and $G_{21}$ are multiplied by the factor $(1-b^{2})/(1+b^{2})$. In the limit
$M\rightarrow\infty$, $b\rightarrow 0$, $Mb^{2}$ finite, the first equation
(\ref{dysongamma1}) still holds, but the second equation (\ref{dysongamma2}) is
replaced by
\begin{eqnarray}
(2\pi E/\delta-4Mb^{2}G_{11})G_{12}=G_{11}\nonumber\\
\mbox{}\times\sum_{n=1}^{N}(-G_{12}+1-2/\Gamma_{n})^{-1}. \label{dysongamma2b}
\end{eqnarray}
The resulting magnetic field dependence of the average density of states is
plotted in Fig.\ \ref{RMTDOSB}, for the case $\Gamma_{n}\equiv 1$ of a
ballistic point contact. The gap closes when $Mb^{2}=N\Gamma/8$. The
corresponding critical flux $\Phi_{c}$ follows from Eq.\ (\ref{crossoverpar}).

\subsection{Broken time-reversal symmetry}
\label{RMTbrokenTRS}

A microscopic suppression of the density of states around $E=0$, on an energy
scale of the order of the level spacing, persists even if time-reversal
symmetry is fully broken. The suppression is a consequence of the level
repulsion between the lowest excitation energy $E_{1}$ and its mirror image
$-E_{1}$, which itself follows from the ${\cal CT}$-antisymmetry
(\ref{CTsymmetry}) of the Hamiltonian.
Because of this mirror symmetry, the effective Hamiltonian ${\cal H}_{\rm eff}$
of the Andreev billiard can be factorized as
\begin{equation}
{\cal H}_{\rm eff}=U\left(\begin{array}{cc}
{\cal E}&0\\0&-{\cal E}\end{array}\right)U^{\dagger},\label{calHUE}
\end{equation}
with $U$ a $2M\times 2M$ unitary matrix and ${\cal E}={\rm
diag}(E_{1},E_{2},\ldots E_{M})$ a diagonal matrix containing the positive
excitation energies.

Altland and Zirnbauer \cite{Alt96} have surmised that an ensemble of Andreev
billiards in a strong magnetic field would have a distribution of Hamiltonians
of the Wigner-Dyson form (\ref{WDPH}), constrained by Eq.\ (\ref{calHUE}). This
constraint changes the Jacobian from the space of matrix elements to the space
of eigenvalues, so that the eigenvalue probability distribution is changed from
the form (\ref{eigenvaluedist}) (with $\beta=2$) into
\begin{equation}
P(\{E_n\})\propto \prod_{i<j} (E_i^2-E_j^2)^2
\prod_k E_k^2 e^{-V(E_{k})-V(-E_{k})}.\label{lag_dist}
\end{equation}
The distribution (\ref{lag_dist}) with $V(E)\propto E^2$ is related to the
Laguerre unitary ensemble (LUE) of RMT \cite{Meh91} by a change of variables.
The average density of states vanishes quadratically near zero energy
\cite{Sle93},
\begin{equation}
\langle\rho(E)\rangle= \frac{2}{\delta}\left(1-\frac{\sin(4\pi E/\delta)}
{4\pi E/\delta}\right).\label{rhoELUE}
\end{equation}
All of this is qualitatively different from the ``folded GUE'' that one would
obtain by simply combining two independent GUE's of electrons and holes
\cite{Bru95}.

A derivation of Altland and Zirnbauer's surmise has been given by Frahm et al.\
\cite{Fra96}, who showed that the LUE for the effective Hamiltonian ${\cal
H}_{\rm eff}$ of the Andreev billiard follows from the GUE for the Hamiltonian
$H$ of the isolated billiard, provided that the coupling to the superconductor
is sufficiently strong. To compute the spectral statistics on the scale of the
level spacing, a non-perturbative technique is needed. This is provided by the
supersymmetric method \cite{Efe97}. (See Ref.\ \cite{Gnu03} for an alternative
approach using quantum graphs.)

The resulting average density of states is \cite{Fra96}
\begin{eqnarray}
\langle\rho(E)\rangle&=&\frac{2}{\delta}
-\frac{\sin(2\pi E/\delta)}{\pi E}\int_0^\infty ds
\ e^{-s}\nonumber\\
&&\mbox{}\times\cos\left(\frac{2\pi E}{\delta}\sqrt{1+\frac{4s}{g_{A}}}
\,\right),\label{rho_result1}\\
g_{A}&=&\sum_{n=1}^{N}\frac{2\Gamma_{n}^{2}}{(2-\Gamma_{n})^{2}}.\label{GAdef}
\end{eqnarray}
The parameter $g_{A}$ is the Andreev conductance of the point contact that
couples the billiard to the superconductor \cite{Bee92}. The Andreev
conductance can be much smaller than the normal-state conductance
$g=\sum_{n=1}^N \Gamma_n$. (Both conductances are in units of $2e^2/h$.) In the
tunneling limit $\Gamma_n\equiv\Gamma\ll 1$ one has $g=N\Gamma$ while
$g_{A}=\frac{1}{2}\,N\Gamma^2$.

\begin{figure}
\centerline{\includegraphics[width=8cm]{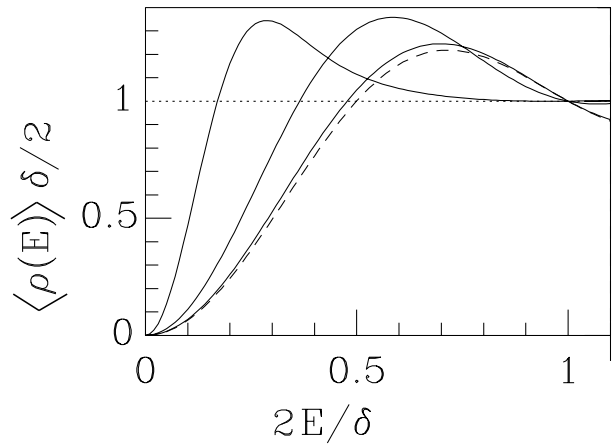}}
\caption{
Density of states of the Andreev billiard in a strong magnetic field for three
different values of the Andreev conductance of the point contact: $g_{A}=
0.4,\ 4,\ 40$. The solid curves are calculated from Eq.\ (\ref{rho_result1}).
The dashed line is the LUE result (\ref{rhoELUE}), corresponding to the limit
$g_{A}\to\infty$. The dotted line is the GUE limit $g_{A}\rightarrow 0$.
Adapted from Ref.\ \cite{Fra96}. \label{RMTDOSC}
}
\end{figure}

Eq.\ (\ref{RMTDOSC}) describes the crossover from the GUE result
$\rho(E)=2/\delta$ for $g_{\rm A}\ll 1$ to the LUE result (\ref{rhoELUE}) for
$g_{\rm A}\gg 1$. The opening of the gap as the coupling to the superconductor
is increased is plotted in Fig.\ \ref{RMTDOSC}. The ${\cal CT}$-antisymmetry
becomes effective at an energy $E$ for $g_{\rm A}\agt E/\delta$. For small
energies $E\ll \delta\,\min (\sqrt{g_{\rm A}},\,1)$ the density of states
vanishes quadratically, regardless of how weak the coupling is.

\subsection{Mesoscopic fluctuations of the gap}
\label{RMTfluct}

The smallest excitation energy $E_{1}$ in the Andreev billiard fluctuates from
one member of the ensemble to the other. Vavilov et al.\ \cite{Vav01} have
surmised that the distribution of these fluctuations is identical upon
rescaling to the known distribution \cite{Tra94} of the lowest eigenvalue in
the Gaussian ensembles of RMT. This surmise was proven using the supersymmetry
technique by Ostrovsky, Skvortsov, and Feigelman \cite{Ost01} and by Lamacraft
and Simons \cite{Lam01}. Rescaling amounts to a change of variables from
$E_{1}$ to $x=(E_{1}-E_{\rm gap})/\Delta_{\rm gap}$, where $E_{\rm gap}$ and
$\Delta_{\rm gap}$ parameterize the square-root dependence (\ref{rhoEneargap})
of the mean density of states near the gap in perturbation theory. The gap
fluctuations are a mesoscopic, rather than a microscopic effect, because the
typical magnitude $\Delta_{\rm gap}\simeq E_{\rm gap}^{1/3}\delta^{2/3}$ of the
fluctuations is $\gg\delta$ for $E_{\rm gap}\gg \delta$. Still, the
fluctuations are small on the scale of the gap itself.

\begin{figure}
\centerline{\includegraphics[width=8cm]{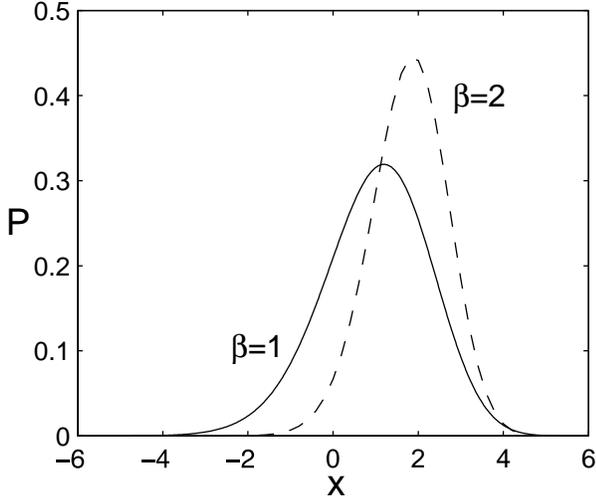}}
\caption{
Probability distribution of the rescaled excitation gap $x =(E_1 - E_{\rm
gap})/\Delta_{\rm gap}$, in the presence [$\beta=1$, Eq.\ (\ref{PE1result})]
and absence [$\beta=2$, Eq.\ (\ref{PE2result})] of time-reversal symmetry.
Adapted from Ref.\ \cite{Vav01}.
\label{PE1}
}
\end{figure}

Following Ref.\ \cite{Vav01}, in zero magnetic field the gap distribution is
obtained by rescaling the GOE result of Tracy and Widom \cite{Tra94},
\begin{eqnarray}
&&P(E_{1})=\frac{d}{dE_{1}}F_{1}\left[(E_{1}-E_{\rm gap})/\Delta_{\rm gap}
\right],\label{PE1result} \\
&&F_{1}(x)=\exp\left( -{\textstyle\frac{1}{2}} \int_{-\infty}^x [q(x')
  + (x - x') q^2(x')] dx' \right).\nonumber\\
&&\label{F1result}
\end{eqnarray}
The function $q(x)$ is the solution of
\begin{equation}
q''(x)=-xq(x)+2q^3(x),\label{qdiffeq}
\end{equation}
with asymptotic behavior $q(x) \to {\rm Ai}(-x)$ as $x\to -\infty$ [${\rm
Ai}(x)$ being the Airy function]. For small $x$ there is a tail of the form
\begin{equation}
  P(x) \approx \frac{1}{4\sqrt{\pi}|x|^{1/4}}
  \exp \left( - {\textstyle\frac{2}{3}}|x|^{3/2} \right), \;\; x \ll -
1.\label{PE1asymp}
\end{equation}
The distribution (\ref{PE1result}) is shown in Fig.\ \ref{PE1} (solid curve).
The mean and standard deviation are
\begin{equation}
\langle E_1 \rangle=E_{\rm gap}+1.21\,\Delta_{\rm gap},\;\;
\langle (E_{1}-\langle E_{1}\rangle)^{2}\rangle^{1/2}=1.27\,\Delta_{\rm
gap}.\label{rmsE1}
\end{equation}

Because the mesoscopic fluctuations in the gap occur on a much smaller energy
scale $\Delta_{\rm gap}$ than $E_{\rm gap}$, there exists a range of magnetic
fields that break time-reversal symmetry of the gap fluctuations without
significantly reducing $E_{\rm gap}$ \cite{Vav01}. In this field range,
specified in Table \ref{fieldrangetable}, the distribution of the lowest
excitation is given by the GUE result \cite{Tra94}
\begin{eqnarray}
&&P(E_{1})=\frac{d}{dE_{1}} F_2[(E_{1} - E_{\rm gap})/\Delta_{\rm gap}],
\label{PE2result}\\
&&F_2(x)\exp \left( - \int_{-\infty}^x (x - x') q^2(x') dx'
\right).\label{F2result}
\end{eqnarray}
This curve is shown dashed in Fig.\ \ref{PE1}. The tail for small $x$ is now
given by
\begin{equation}
P(x)  \approx \frac{1}{8\pi |x|} \exp\left(-{\textstyle\frac{4}{3}}|x|^{3/2}
\right), \;\; x \ll -1.\label{PE1asymp2}
\end{equation}

\begin{table}
\centerline{
\begin{tabular}{ c  c  c }
  & Energy scale   & Flux scale\\
\hline
Bulk statistics & $\delta$ & $(h/e) \tau_{\rm erg}^{1/2}
\delta^{1/2}/\hbar^{1/2}$ \\
Edge statistics & $E_{\rm gap}^{1/3} \delta^{2/3}$ & $(h/e)\tau_{\rm erg}^{1/2}
  \delta^{1/6} E_{\rm gap}^{1/3}/\hbar^{1/2}$ \\
Gap size & $E_{\rm gap}$ & $(h/e) \tau_{\rm erg}^{1/2} E_{\rm
gap}^{1/2}/\hbar^{1/2}$ \\
\end{tabular}
}
\caption{
Characteristic energy and magnetic flux scales for the
spectral statistics in the bulk and at the edge of the spectrum
and for the size of the gap. The $\beta=2$ distribution (\ref{PE2result})
applies to the flux range $(h/e)\tau_{\rm erg}^{1/2}\delta^{1/6} E_{\rm
gap}^{1/3}/\hbar^{1/2}\ll\Phi\ll (h/e) \tau_{\rm erg}^{1/2} E_{\rm
gap}^{1/2}/\hbar^{1/2}$.
\label{fieldrangetable}
}
\end{table}

\begin{figure}
\centerline{\includegraphics[width=8cm]{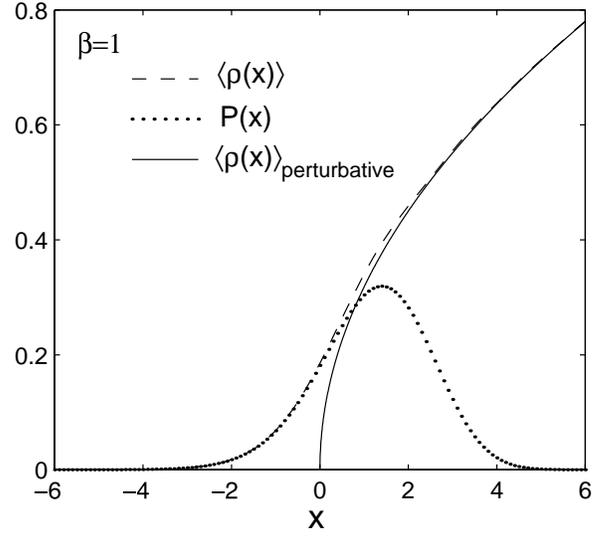}}
\caption{
Ensemble averaged density of states $\langle\rho\rangle$ together with the
probability
distribution $P$ of the excitation gap, as a function of the rescaled energy
$x=(E-E_{\rm gap})/\Delta_{\rm gap}$. The dotted and dashed curves are the
universal results (\ref{PE1result}) and (\ref{rhox}) of RMT in the presence of
time-reversal symmetry ($\beta=1$). The solid curve is the mean density of
states (\ref{rhoEneargap}) in perturbation theory. Adapted from Ref.\
\cite{Vav01}.
\label{RMTDOSD}
}
\end{figure}

The mesoscopic gap fluctuations induce a tail in the ensemble averaged density
of states $\langle\rho(E)\rangle$ for $E<E_{\rm gap}$. In the same rescaled
variable $x$ the tail is given by \cite{Vav01}
\begin{eqnarray}
\langle \rho(x) \rangle&=&-x{\rm Ai}^2(x)+[{\rm Ai}'(x)]^2\nonumber\\
&&\mbox{}+{\textstyle\frac{1}{2}} \delta_{\beta,1} {\rm
Ai}(x)[1-\int_x^{\infty} {\rm Ai}(y)dy ].\label{rhox}
\end{eqnarray}
Asymptotically, $\langle \rho(x) \rangle \propto  \exp(-\frac{2}{3}\beta
|x|^{3/2})$ for $x\ll -1$. The tail in $\langle\rho(E)\rangle$ is the same as
the tail in $P(E)$, as it should be, since both tails are due to the lowest
eigenvalue. In Fig.\ \ref{RMTDOSD} we compare these two functions in zero
magnetic field, together with the square-root density of states from
perturbation theory.

\begin{figure}
\centerline{\includegraphics[width=8cm]{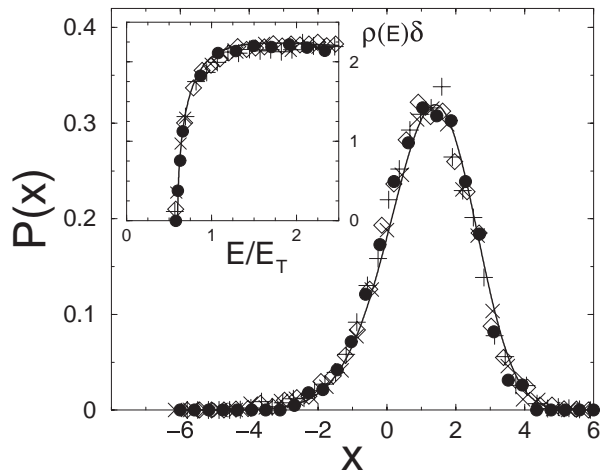}}
\caption{
Main plot: Gap distribution for the Andreev kicked rotator with parameters
$M=2\pi/\delta=8192$, kicking strength $K=45$, and $M/N=\tau_{\rm dwell}=10$
($\diamond$), 20 ($\bullet$), 40 ($+$), and 50 ($\times$). There is no magnetic
field. The solid line is the RMT prediction (\ref{PE1result}). Inset: Average
density of states for the same system. The solid line is the RMT prediction
(\ref{rhoEballisticc}). (Deviations from perturbation theory are not visible on
the scale of the inset.) Adapted from Ref.\ \cite{Jac03}.
\label{gapflucttest}
}
\end{figure}

A numerical simulation of the stroboscopic model of Sec.\ \ref{strobomodel}
provides a test of these predictions \cite{Jac03}. Results are shown in Fig.\
\ref{gapflucttest}, for the case $\beta=1$ and deep in the chaotic regime
(kicking strength $K\gg 1$). The agreement with RMT is very good --- without
any adjustable parameters.

\subsection{Coulomb blockade}

Coulomb interactions between electron and hole quasiparticles break the
charge-conjugation invariance (\ref{CTsymmetry}) of the Hamiltonian. Since
Andreev reflection changes the charge on the billiard by $2e$, this scattering
process becomes energetically unfavorable if the charging energy $E_{C}$
exceeds the superconducting condensation energy (Josephson energy) $E_{J}$. For
$E_{C}\agt E_{J}$ one obtains the Coulomb blockade of the proximity effect
studied by Ostrovsky, Skvortsov, and Feigelman \cite{Ost04}.

The charging energy $E_{C}=e^{2}/2C$ is determined by the capacitance $C$ of
the billiard. The Josephson energy is determined by the change in free energy
of the billiard resulting from the coupling to the superconductor,
\begin{equation}
E_{J}=-\int_{0}^{\infty}[\rho(E)-2/\delta]\,EdE.\label{EJdef}
\end{equation}
The discrete spectrum $E<E_{\rm gap}$ contributes an amount of order $E_{\rm
gap}^{2}/\delta$ to $E_{J}$. In the continuous spectrum $E>E_{\rm gap}$ the
density of states $\rho(E)$, calculated by RMT, decays $\propto 1/E^{2}$ to its
asymptotic value $2/\delta$. This leads to a logarithmic divergence of the
Josephson energy \cite{Bro97a,Asl68}, with a cutoff set by
$\min(\Delta,\hbar/\tau_{\rm erg})$:
\begin{equation}
E_{J}=\frac{E_{\rm gap}^{2}}{\delta}\ln\left(\frac{\min(\Delta,\hbar/\tau_{\rm
erg})}{E_{\rm gap}}\right).\label{EJresult}
\end{equation}

\begin{figure}
\centerline{\includegraphics[width=8cm]{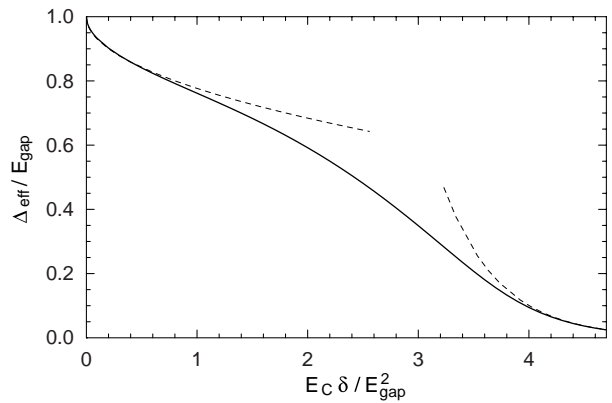}}
\caption{
Suppression due to Coulomb interactions of the gap $\Delta_{\rm eff}$ in the
density of states of an Andreev billiard coupled by a tunnel junction to a
superconductor, relative to the noninteracting gap $E_{\rm
gap}=N\Gamma\delta/4\pi$ (with $\Gamma\ll 1\ll N\Gamma$). The plot is for the
case $\Delta=e^{5}E_{\rm gap}\ll\hbar/\tau_{\rm erg}$. The dashed lines are the
asymptotes (\ref{Etilde1}) and (\ref{Etilde2}). Adapted from Ref. \cite{Ost04}.
\label{Coulombgap1}
}
\end{figure}

The suppression of the excitation gap with increasing $E_{C}$ is plotted in
Fig.\ \ref{Coulombgap1}, for the case $\Gamma\ll 1$, $\Delta\ll\hbar/\tau_{\rm
erg}$ \cite{Ost04}. The initial decay is a square root,
\begin{equation}
1-\Delta_{\rm eff}/E_{\rm
gap}={\textstyle\frac{1}{2}}\left(\frac{E_{C}\delta}{E_{\rm
gap}^{2}\ln(2\Delta/E_{\rm gap})}\right)^{1/2}\ll 1,\label{Etilde1}
\end{equation}
and the final decay is exponential,
\begin{equation}
\Delta_{\rm eff}/\Delta=2\exp(-2E_{C}\delta/E_{\rm gap}^{2})\ll
1.\label{Etilde2}
\end{equation}
Here $\Delta_{\rm eff}$ refers to the gap in the presence of Coulomb
interactions and $E_{\rm gap}=N\Gamma\delta/4\pi$ is the noninteracting value
(\ref{rhoEtunnel}).

\begin{figure}
\centerline{\includegraphics[width=8cm]{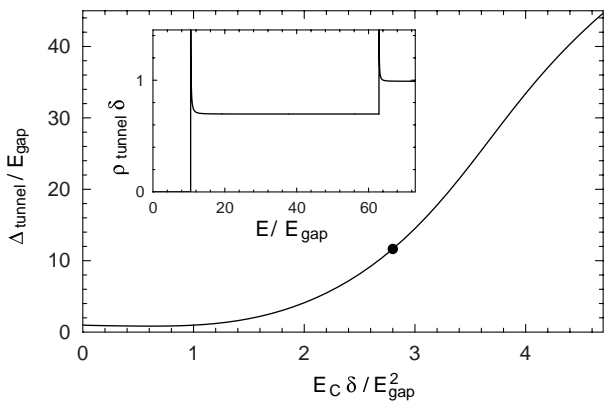}}
\caption{
Main plot: gap $\Delta_{\rm tunnel}$ in the tunneling density of states as a
function of the charging energy (for $\Delta=e^{5}E_{\rm gap}$ and
$N\Gamma=40\,\pi$). The initial decay (barely visible on the scale of the plot)
follows Eq.\ (\ref{Etilde1}) and crosses over to an increase ($\Delta_{\rm
tunnel}\rightarrow E_{C}$). Inset: tunneling density of states at
$E_{C}\delta/E_{\rm gap}^{2}=2.8$ (corresponding to the dot in the main plot).
Adapted from Ref. \cite{Ost04}.
\label{Coulombgap2}
}
\end{figure}

The gap $\Delta_{\rm eff}$ governs the thermodynamic properties of the Andreev
billiard, most importantly the critical current. It is not, however, the
relevant energy scale for transport properties. Injection of charge into the
billiard via a separate tunnel contact measures the tunneling density of states
$\rho_{\rm tunnel}$, which differs in the presence of Coulomb interactions from
the thermodynamic density of states $\rho$ considered so far. The gap
$\Delta_{\rm tunnel}$ in $\rho_{\rm tunnel}$ crosses over from the proximity
gap $E_{\rm gap}$ when $E_{C}\ll E_{J}$ to the Coulomb gap $E_{C}$ when
$E_{C}\gg E_{J}$, see Fig.\ \ref{Coulombgap2}. The single peak in $\rho_{\rm
tunnel}$ at $\Delta_{\rm tunnel}$ splits into two peaks when $E_{C}$ and
$E_{J}$ are of comparable magnitude \cite{Ost04}. This peak splitting happens
because two states of charge $+e$ and $-e$ having the same charging energy are
mixed by Andreev reflection into symmetric and antisymmetric linear
combinations with a slightly different energy.

\section{Quasiclassical theory}
\label{quasiclassics}

It was noticed by Kosztin, Maslov, and Goldbart \cite{Kos95} that the classical
dynamics at the Fermi energy in an Andreev billiard is {\em integrable\/} ---
even if the dynamics in the isolated billiard is chaotic. Andreev reflection
suppresses chaotic dynamics because it introduces a periodicity into the
orbits: The trajectory of an electron is retraced by the Andreev reflected
hole. At the Fermi energy the hole is precisely the time reverse of the
electron, so that the motion is strictly periodic. For finite excitation energy
or in a non-zero magnetic field the electron and the hole follow slightly
different trajectories, so the orbit does not quite close and drifts around in
phase space \cite{Kos95,Shy98,Wie02,Ada02,Cse03}.

The near-periodicity of the orbits implies the existence of an adiabatic
invariant. Quantization of this invariant leads to the quasiclassical theory of
Silvestrov et al.\ \cite{Sil03}.

\subsection{Adiabatic quantization}
\label{adiabaticquant}

\begin{figure}
\centerline{\includegraphics[width=8cm]{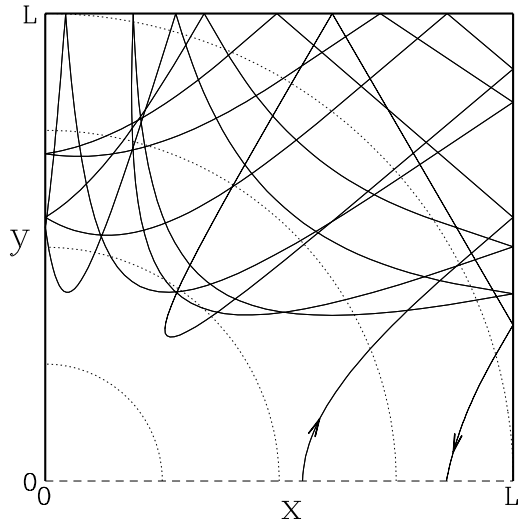}}
\caption{
Classical trajectory in an Andreev billiard. Particles in a two-dimensional
electron gas are deflected by an electrostatic potential. (The dotted circles
are equipotentials.) There is specular reflection at the boundaries with an
insulator (thick solid lines) and Andreev reflection at the boundary with a
superconductor (dashed line). The trajectory follows the motion between two
Andreev reflections of an electron near the Fermi energy. The Andreev reflected
hole retraces this trajectory in opposite direction. From Ref.\ \cite{Sil03}.
\label{paden}
}
\end{figure}

\begin{figure}
\centerline{\includegraphics[width=8cm]{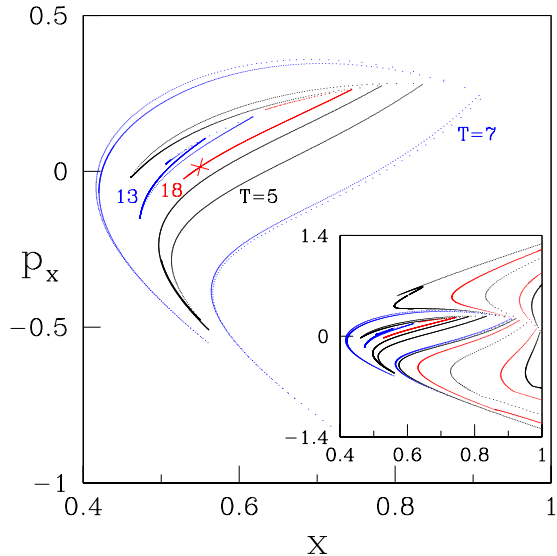}}
\caption{
Poincar\'{e} map for the Andreev billiard of Fig.\ \ref{paden}. Each dot
represents a starting point of an electron trajectory, at position $x$ and with
tangential momentum $p_{x}$ (in dimensionless units). The inset shows the full
surface of section, while the main plot is an enlargement of the central
region. The drifting nearly periodic motion follows contours of constant time
$T$ between Andreev reflections. The cross marks the starting point of the
trajectory shown in the previous figure. From Ref.\ \cite{Sil03}.
\label{phasespace}
}
\end{figure}

Figs.\ \ref{paden} and \ref{phasespace} illustrate the nearly periodic motion
in a particular Andreev billiard. Fig.\ \ref{paden} shows a trajectory in real
space while Fig.\ \ref{phasespace} is a section of phase space at the interface
with the superconductor ($y=0$). The tangential component $p_{x}$ of the
electron momentum is plotted as a function of the coordinate $x$ along the
interface. Each point in this Poincar\'{e} map corresponds to one collision of
an electron with the interface. (The collisions of holes are not plotted.) The
electron is retroreflected as a hole with the same $p_{x}$. At the Fermi level
($E=0$) the component $p_{y}$ is also the same, and so the hole retraces the
path of the electron (the hole velocity being opposite to its momentum). The
Poincar\'{e} map would then consist of a single point. At non-zero excitation
energy $E$ the retroreflection occurs with a slight change in $p_{y}$, because
of the difference $2E$ in the kinetic energy of electrons (at energy $E_{F}+E$)
and holes (at energy $E_{F}-E$).

The resulting slow drift of the periodic trajectory traces out a contour in the
surface of section. These are {\em isochronous\/} contours \cite{Sil03},
meaning that the time $T$ between Andreev reflections is the same for each
point $x,p_{x}$ on the contour . The adiabatic invariance of $T$ follows from
the adiabatic invariance of the action integral $I$ over the nearly periodic
motion from electron to hole and back to electron:
\begin{equation}
I=\oint pdq=2E T.\label{Idef}
\end{equation}
Since $E$ is a constant of the motion, adiabatic invariance of $I$ implies
adiabatic invariance of the time $T$ between Andreev reflections.

Adiabatic invariance is defined in the limit $E\rightarrow 0$ and is therefore
distinct from invariance in the sense of Kolmogorov-Arnold-Moser~(KAM)
\cite{Gut90}, which would require a critical $E^*$ such that a contour is
exactly invariant for $E<E^*$. There is numerical evidence \cite{Kos95} that
the KAM theorem does not apply to a chaotic Andreev billiard.

It is evident from Fig.\ \ref{phasespace} that contours of large $T$ enclose a
very small area in a chaotic system. To estimate the area, it is convenient to
measure $x$ in units of the width $W$ of the constriction to the
superconductor. Similarly, $p_{x}$ is conveniently measured in units of the
range $\Delta p$ of transverse momenta inside the constriction.\footnote{
We consider in this estimate the symmetric case $W/L\simeq \Delta p/p_{F}\ll
1$, typical for the smooth confining potential of Fig.\ \protect\ref{paden}. In
the asymmetric case $W/L\ll\Delta p/p_{F}\simeq 1$, typical for the computer
simulations using the kicked rotator, the maximal area $A_{\rm max}$ is smaller
by a factor $W/L$, cf.\ Ref.\ \protect\cite{Sil03b}. Consequently, the factor
$\ln N$ in Eq.\ (\protect\ref{epsilon00}) should be replaced by $\ln(NW/L)$.}
The highly elongated shape evident in Fig.\ \ref{phasespace} is a consequence
of the exponential divergence in time of nearby trajectories, characteristic of
chaotic dynamics. The rate of divergence is the Lyapunov exponent $\alpha$.
Since the Hamiltonian flow is area preserving, a stretching
$\ell_{+}(t)=\ell_{+}(0)e^{\alpha t}$ of the dimension in one direction needs
to be compensated by a squeezing $\ell_{-}(t)=\ell_{-}(0)e^{-\alpha t}$ of the
dimension in the other direction. The area $A\simeq\ell_{+}\ell_{-}$ is then
time independent. Initially, $\ell_{\pm}(0)<1$. The constriction at the
superconductor acts as a bottleneck, enforcing $\ell_{\pm}(T)<1$. These two
inequalities imply $\ell_{+}(t)<e^{\alpha(t-T)}$, $\ell_{-}<e^{-\alpha t}$. The
enclosed area, therefore, has upper bound
\begin{equation}
A_{\rm max}\simeq W\Delta p\,e^{-\alpha T}\simeq\hbar Ne^{-\alpha
T},\label{Amax}
\end{equation}
where $N\simeq W\Delta p/\hbar\gg 1$ is the number of channels in the point
contact.

The two invariants $E$ and $T$ define a two-dimensional torus in the
four-dimensional phase space. Quantization of this adiabatically invariant
torus proceeds following Einstein-Brillouin-Keller \cite{Gut90,Dun02}, by
quantizing the area
\begin{equation}
\oint pdq=2\pi\hbar(m+\nu/4),\;\;m=0,1,2,\ldots\label{EBK}
\end{equation}
enclosed by each of the two topologically independent contours on the torus.
Eq.\ (\ref{EBK}) ensures that the wave functions are single valued. The integer
$\nu$ counts the number of caustics (Maslov index) and in this case should also
include the number of Andreev reflections.

The first contour follows the quasiperiodic orbit of Eq.\ (\ref{Idef}), leading
to
\begin{equation}
ET=(m+{\textstyle\frac{1}{2}})\pi\hbar,
\;\;m=0,1,2,\ldots\label{epsilonTquant}
\end{equation}
The quantization condition (\ref{epsilonTquant}) is sufficient to determine the
smoothed (or ensemble averaged) density of states
\begin{equation}
\langle\rho(E)\rangle=N\int_{0}^{\infty}dT\,P(T)
\sum_{m=0}^{\infty}\delta\bigl(E-
(m+{\textstyle\frac{1}{2}})\pi\hbar/T\bigr), \label{rhoBS1}
\end{equation}
using the classical probability distribution $P(T)$ for the time between
Andreev reflections. (The distribution $P(T)$ is defined with a uniform measure
in the surface of section $(x,p_{x})$ at the interface with the
superconductor.)

Eq.\ (\ref{rhoBS1}) is the ``Bohr-Sommerfeld rule'' of Melsen et al.\
\cite{Mel96}. It generalizes the familiar Bohr-Sommerfeld quantization rule for
translationally invariant geometries [cf.\ Eq.\ (\ref{E0ppar})] to arbitrary
geometries. The quantization rule refers to classical periodic motion with
period $2T$ and phase increment per period of $2ET/\hbar-\pi$, consisting of a
part $2ET/\hbar$ because of the energy difference $2E$ between electron and
hole, plus a phase shift of $-\pi$ from two Andreev reflections. If $E$ is not
$\ll\Delta$, this latter phase shift should be replaced by
$-2\arccos(E/\Delta)$ \cite{Cse02,Cse02b,Cse04}, cf.\ Eq.\ (\ref{alphadef}). In
the presence of a magnetic field an extra phase increment proportional to the
enclosed flux should be included \cite{Ihr01}. Eq.\ (\ref{rhoBS1}) can also be
derived from the Eilenberger equation for the quasiclassical Green function
\cite{Lod98}.

To find the location of individual energy levels a second quantization
condition is needed \cite{Sil03}. It is provided by the area $\oint_{T}
p_{x}dx$ enclosed by the isochronous contours,
\begin{equation}
\oint_{T}p_{x}dx=2\pi\hbar(n+\nu/4),\;\;n=0,1,2,\ldots\label{Tquant}
\end{equation}
Eq.\ (\ref{Tquant}) amounts to a quantization of the period $T$, which together
with Eq.\ (\ref{epsilonTquant}) leads to a quantization of $E$. For
each $T_{n}$ there is a ladder of Andreev levels
$E_{nm}=(m+{\textstyle\frac{1}{2}})\pi\hbar/T_{n}$.

While the classical $T$ can become arbitrarily large, the quantized $T_{n}$ has
a cutoff. The cutoff follows from the maximal area (\ref{Amax}) enclosed by an
isochronous contour. Since Eq.\ (\ref{Tquant}) requires $A_{\rm max}\agt h$,
the longest quantized period is $T_{0}=\alpha^{-1}[\ln N+{\cal O}(1)]$. The
lowest Andreev level associated with an adiabatically invariant torus is
therefore
\begin{equation}
E_{00}=\frac{\pi\hbar}{2T_{0}}=\frac{\pi\hbar\alpha}{2\ln
N}.\label{epsilon00}
\end{equation}
The time scale $T_{0}\propto|\ln\hbar|$ is the Ehrenfest time $\tau_{E}$ of the
Andreev billiard, to which we will return in Sec.\ \ref{clqcrossover}.

The range of validity of adiabatic quantization is determined by the
requirement that the drift $\delta x$, $\delta p_{x}$ upon one iteration of the
Poincar\'{e} map should be small compared to the characteristic values
$W,p_{F}$. An estimate is \cite{Sil03}
\begin{equation}
\frac{\delta x}{W}\simeq\frac{\delta
p_{x}}{p_{F}}
\simeq \frac{E_{nm}}{\hbar\alpha N}e^{\alpha T_{n}}
\simeq (m+{\textstyle \frac{1}{2}})\frac{e^{-\alpha (T_0-T_{n})}}{\alpha
T_{n}}.
\label{deltaestimate}
\end{equation}
For low-lying levels ($m\sim 1$) the dimensionless drift is $\ll 1$ for
$T_n<T_0$. Even for $T_n=T_0$ one has $\delta x/W\simeq 1/\ln N\ll 1$.

\subsection{Integrable dynamics}
\label{integrabledyn}

Unlike RMT, the quasiclassical theory is not restricted to systems with a
chaotic classical dynamics. Melsen et al.\ \cite{Mel96,Mel97} have used the
Bohr-Sommerfeld rule (\ref{rhoBS1}) to argue that Andreev billiards with an
integrable classical dynamics have a smoothly vanishing density of states ---
without an actual excitation gap. The presence or absence of an excitation gap
is therefore a ``quantum signature of chaos''. This is a unique property of
Andreev billiards. In normal, not-superconducting billiards, it is impossible
to distinguish chaotic from integrable dynamics by looking at the density of
states. One needs to measure density-density correlation functions for that
purpose \cite{Haa01}.

\begin{figure}
\centerline{\includegraphics[width=8cm]{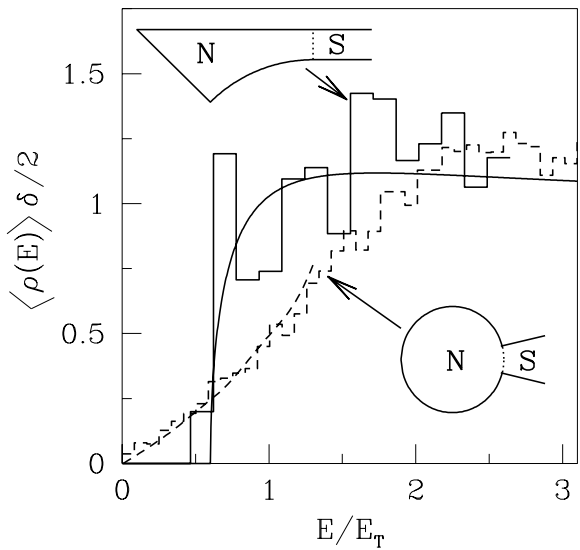}}
\caption{
Histograms: smoothed density of states of a billiard coupled by a ballistic
$N$-mode lead to a superconductor, determined by Eq.\ (\ref{discreteE}) and
averaged over a range of Fermi energies at fixed $N$. The scattering matrix is
computed numerically by matching wave functions in the billiard to transverse
modes in the lead. A chaotic Sinai billiard (top inset, solid histogram,
$N=20$) is contrasted with an integrable circular billiard (bottom inset,
dashed histogram, $N=30$). The solid curve is the prediction
(\ref{rhoEballisticc}) from RMT for a chaotic system and the dashed curve is
the Bohr-Sommerfeld result (\ref{rhoBS1}), with dwell time distribution $P(T)$
calculated from classical trajectories in the circular billiard. Adapted from
Ref.\ \cite{Mel97}.
\label{circlebilliard}
}
\end{figure}

The difference between chaotic and integrable Andreev billiards is illustrated
in Fig.\ \ref{circlebilliard}. As expected, the chaotic Sinai billiard follows
closely the prediction from RMT. (The agreement is less precise than for the
kicked rotator of Fig.\ \ref{gapflucttest}, because the number of modes $N=20$
is necessarily much smaller in this simulation.) The density of states of the
integrable circular billiard is suppressed on the same mesoscopic energy scale
$E_{T}$ as the chaotic billiard, but the suppression is smooth rather than
abrupt. Any remaining gap is microscopic, on the scale of the level spacing,
and therefore invisible in the smoothed density of states.

That the absence of an excitation gap is generic for integrable billiards can
be understood from the Bohr-Sommerfeld rule \cite{Mel96}. Generically, an
integrable billiard has a power-law distribution of dwell times, $P(T)\propto
T^{-p}$ for $T\rightarrow\infty$, with $p\approx 3$ \cite{Bau90,Bar93}. Eq.\
(\ref{rhoBS1}) then implies a power-law density of states,
$\langle\rho(E)\rangle\propto E^{p-2}$ for $E\rightarrow 0$. The value $p=3$
corresponds to a linearly vanishing density of states. An analytical
calculation \cite{Kor03} of $P(T)$ for a rectangular billiard gives the
long-time limit $P(T)\propto T^{-3}\ln T$, corresponding to the low-energy
asymptote $\langle\rho(E)\rangle\propto E\ln(E_{T}/E)$. The weak logarithmic
correction to the linear density of states is consistent with exact quantum
mechanical calculations \cite{Mel96,Ihr01}.

\subsection{Chaotic dynamics}
\label{chaoticdyn}

\begin{figure}
\centerline{\includegraphics[width=8cm]{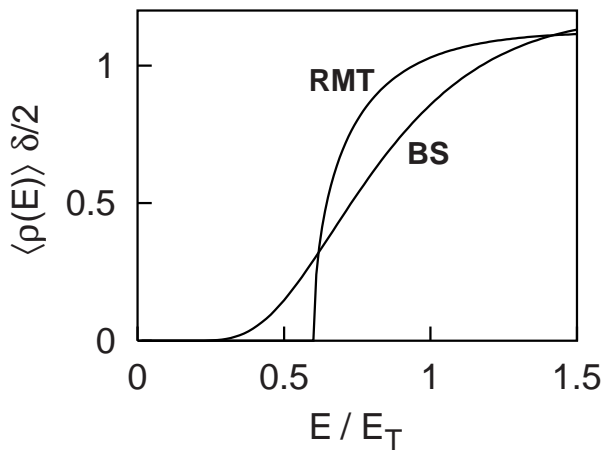}}
\caption{
Comparison of the smoothed density of states in a chaotic Andreev billiard as
it follows from RMT (Eq.\ (\ref{rhoEballisticc}), with a hard gap) and as it
follows from the Bohr-Sommerfeld (BS) rule (Eq.\ (\ref{chaoticdos}), without a
hard gap). These are the two limiting distributions when the Ehrenfest time
$\tau_{E}$ is, respectively, much smaller or much larger than the mean dwell
time $\tau_{\rm dwell}$.
\label{twoDOS}
}
\end{figure}

A chaotic billiard has an exponential dwell time distribution, $P(T)\propto
e^{-T/\tau_{\rm dwell}}$, instead of a power law \cite{Bau90}. (The mean dwell
time is $\tau_{\rm dwell}=2\pi\hbar/N\delta\equiv \hbar/2E_{T}$.) Substitution
into the Bohr-Sommerfeld rule (\ref{rhoBS1}) gives the density of states
\cite{Sch99}
\begin{equation}
\langle\rho(E)\rangle=\frac{2}{\delta}\,\frac{(\pi E_{T}/E)^{2}\cosh (\pi
E_{T}/E)}{\sinh^{2} (\pi E_{T}/E)},\label{chaoticdos}
\end{equation}
which vanishes $\propto e^{-\pi E_{T}/E}$ as $E\rightarrow 0$. This is a much
more rapid decay than for integrable systems, but not quite the hard gap
predicted by RMT \cite{Mel96}. The two densities of states are compared in
Fig.\ \ref{twoDOS}.

When the qualitative difference between the random-matrix and Bohr-Sommerfeld
theories was discovered \cite{Mel96}, it was believed to be a short-coming of
the quasiclassical approximation underlying the latter theory. Lodder and
Nazarov \cite{Lod98} realized that the two theoretical predictions are actually
both correct, in different limits. As the ratio $\tau_{E}/\tau_{\rm dwell}$ of
Ehrenfest time and dwell time is increased, the density of states crosses over
from the RMT form (\ref{rhoEballisticc}) to the Bohr-Sommerfeld form
(\ref{chaoticdos}). We investigate this crossover in the following section.

\section{Quantum-to-classical crossover}
\label{clqcrossover}

\subsection{Thouless versus Ehrenfest}
\label{TvsE}

\begin{figure}
\centerline{\includegraphics[width=6cm]{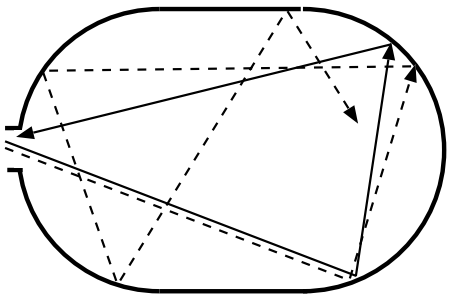}}
\caption{
Two trajectories entering a chaotic billiard at a small separation $\delta
x(0)$ diverge exponentially in time, $\delta x(t)=\delta x(0)e^{\alpha t}$. The
rate of divergence $\alpha$ is the Lyapunov exponent. An initial microscopic
separation $\lambda_{F}$ becomes macroscopic at the Ehrenfest time
$\tau_{E}=\alpha^{-1}\ln(L^{\ast}/\lambda_{F})$. The macroscopic length
$L^{\ast}$ is determined by the size and shape of the billiard. The Ehrenfest
time depends logarithmically on Planck's constant:
$\tau_{E}=\alpha^{-1}\ln(S_{\rm cl}/h)$, with $S_{\rm cl}=mv_{F}L^{\ast}$ the
characteristic classical action. The evolution of a quantum mechanical wave
packet is well described by a classical trajectory only for times less than
$\tau_{E}$.
\label{andreevbilliard}
}
\end{figure}

According to Ehrenfest's theorem, the propagation of a quantum mechanical
wave packet is described for short times by classical equations of
motion. The time scale at which this correspondence between quantum
and classical dynamics breaks down in a chaotic system is called
the Ehrenfest time $\tau_{E}$ \cite{Ber78}.\footnote{The name ``Ehrenfest
time'' was coined in Ref.\ \cite{Chi88}.} As explained in
Fig.\ \ref{andreevbilliard}, it depends logarithmically on Planck's
constant: $\tau_{E}=\alpha^{-1}\ln(S_{\rm cl}/h)$, with $S_{\rm cl}$
the characteristic classical action of the dynamical system and $\alpha$
the Lyapunov exponent.

This logarithmic $h$-dependence distinguishes
the Ehrenfest time from other characteristic time scales of a chaotic
system, which are either $h$-independent (dwell time, ergodic time)
or algebraically dependent on $h$ (Heisenberg time $\propto 1/\delta$).
That the quasiclassical theory of superconductivity breaks down on time scales
greater than $\tau_{E}$ was noticed already in 1968 by Larkin and Ovchinnikov
\cite{Lar68}.

The choice of $S_{\rm cl}$ depends on the physical quantity which one is
studying. For the density of states of the Andreev billiard (area $A$, opening
of width $W\ll A^{1/2}$, range of transverse momenta $\Delta p\simeq p_{F}$
inside the constriction) the characteristic classical action is\footnote{
The simpler expression $S_{\rm cl}=mv_{F}W$ of Ref.\ \protect\cite{Sil03}
applies to the symmetric case $W/A^{1/2}\simeq\Delta p/p_{F}\ll 1$.}
$S_{\rm cl}=mv_{F}W^{2}/A^{1/2}$ \cite{Vav03}. The Ehrenfest time then takes
the form
\begin{equation}
\tau_{E}=\alpha^{-1}[\ln(N^{2}/M)+{\cal O}(1)].\label{tauEdef}
\end{equation}
Here $M=k_{F}A^{1/2}/\pi$ and $N=k_{F}W/\pi$ are, respectively, the number of
modes in a cross-section of the billiard and in the point contact. Eq.\
(\ref{tauEdef}) holds for $N\agt\sqrt{M}$. For $N\alt\sqrt{M}$ the Ehrenfest
time may be set to zero, because the wave packet then spreads over the entire
billiard within the ergodic time \cite{Sil03b}.

Chaotic dynamics requires $\alpha^{-1}\ll\tau_{\rm dwell}$. The relative
magnitude of $\tau_{E}$ and $\tau_{\rm dwell}$ thus depends on whether the
ratio $N^{2}/M$ is large or small compared to the exponentially large number
$e^{\alpha\tau_{\rm dwell}}$.

The result of RMT \cite{Mel96}, cf.\ Sec.\ \ref{RMTexcgap}, is that the
excitation gap in an Andreev billiard is of the order of the Thouless energy
$E_{T}\simeq\hbar/\tau_{\rm dwell}$. It was realized by Lodder and Nazarov
\cite{Lod98} that this result requires $\tau_{E}\ll\tau_{\rm dwell}$. More
generally, the excitation gap $E_{\rm gap}\simeq{\rm
min}\,(E_{T},\hbar/\tau_{E})$ is determined by the smallest of the Thouless and
Ehrenfest energy scales. The Bohr-Sommerfeld theory \cite{Mel96}, cf.\ Sec.\
\ref{chaoticdyn}, holds in the limit $\tau_{E}\rightarrow\infty$ and therefore
produces a gapless density of states.

\subsection{Effective RMT}
\label{EffRMT}

A phenomenological description of the crossover from the Thouless to the
Ehrenfest
regime is provided by the ``effective RMT'' of Silvestrov et al.\ \cite{Sil03}.
As described in Sec.\ \ref{adiabaticquant}, the quasiclassical adiabatic
quantization allows to quantize only the trajectories with periods
$T\leq T_{0}\equiv\tau_{E}$. The excitation gap of the Andreev billiard is
determined by the part of phase space with periods longer than $\tau_{E}$.
Effective RMT is based on the hypothesis
that this part of phase space can be quantized by a scattering matrix $S_{\rm
eff}$ in the circular ensemble of RMT, with a reduced dimensionality
\begin{equation}
N_{\rm eff}=N\int_{\tau_{E}}^{\infty}P(T)\,dT= Ne^{-\tau_{E}/\tau_{\rm dwell}}.
\label{Neffdef}
\end{equation}

\begin{figure}
\centerline{\includegraphics[width=6cm]{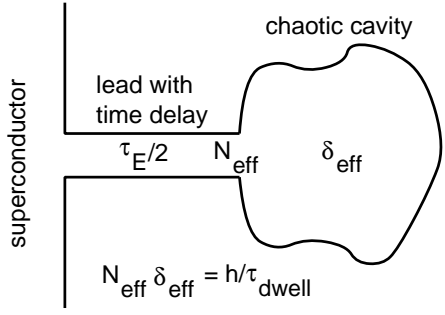}}
\caption{Pictorial representation of the effective RMT of an Andreev billiard.
The part of phase space with time $T>\tau_{E}$ between Andreev reflections is
represented by a chaotic cavity (mean level spacing $\delta_{\rm eff}$),
connected to the superconductor by a long lead ($N_{\rm eff}$ propagating
modes, one-way delay time $\tau_{E}/2$ for each mode). Between two Andreev
reflections an electron or hole spends, on average, a time $\tau_{E}$ in the
lead and a time $\tau_{\rm dwell}$ in the cavity. The scattering matrix of lead
plus cavity is $\exp(iE\tau_{E}/\hbar)S_{0}(E)$, with $S_{0}(E)$ distributed
according to the circular ensemble of RMT (with effective parameters $N_{\rm
eff}$, $\delta_{\rm eff}$). The complete excitation spectrum of the Andreev
billiard consists of the levels of the effective RMT (periods $>\tau_{E}$) plus
the levels obtained by adiabatic quantization (periods $<\tau_{E}$).
\label{cavitylead}
}
\end{figure}

The energy dependence of $S_{\rm eff}(E)$ is that of a chaotic cavity with mean
level spacing $\delta_{\rm eff}$, coupled to the superconductor by a long lead
with $N_{\rm eff}$ propagating modes. (See Fig.\ \ref{cavitylead}.) The lead
introduces a mode-independent delay time $\tau_{E}$ between Andreev
reflections, to ensure that $P(T)$ is cut off for $T<\tau_{E}$. Because $P(T)$
is exponential $\propto\exp(-T/\tau_{\rm dwell})$, the mean time $\langle
T\rangle_{\ast}$ between Andreev reflections in the accessible part of phase
space is simply $\tau_{E}+\tau_{\rm dwell}$. The effective level spacing in the
chaotic cavity by itself (without the lead) is then determined by
\begin{equation}
\frac{2\pi\hbar}{N_{\rm eff}\delta_{\rm eff}}=\langle
T\rangle_{\ast}-\tau_{E}=\tau_{\rm dwell}.\label{deltaeffdef}
\end{equation}

It is convenient to separate the energy dependence due to the lead from that
due to the cavity, by writing $S_{\rm eff}(E)=\exp(iE\tau_{E}/\hbar)S_{0}(E)$,
where $S_{0}(E)$ represents only the cavity and has an energy dependence of the
usual RMT form (\ref{SWHformula}) --- with effective parameters $N_{\rm eff}$
and $\delta_{\rm eff}$. The determinant equation (\ref{discreteE}) for the
excitation spectrum then takes the form
\begin{equation}
{\rm
Det}\,\left[1-\alpha(E)^{2}e^{2iE\tau_{E}/\hbar}
S_{0}(E)S_{0}(-E)^{\ast}\right]=0.\label{discreteE2}
\end{equation}
We can safely replace $\alpha(E)\equiv\exp[-i\arccos(E/\Delta)]\rightarrow-i$
(since $E\ll\Delta$), but the energy dependence of the phase factor
$e^{2iE\tau_{E}/\hbar}$ can not be omitted.

\begin{figure}
\centerline{\includegraphics[width=8cm]{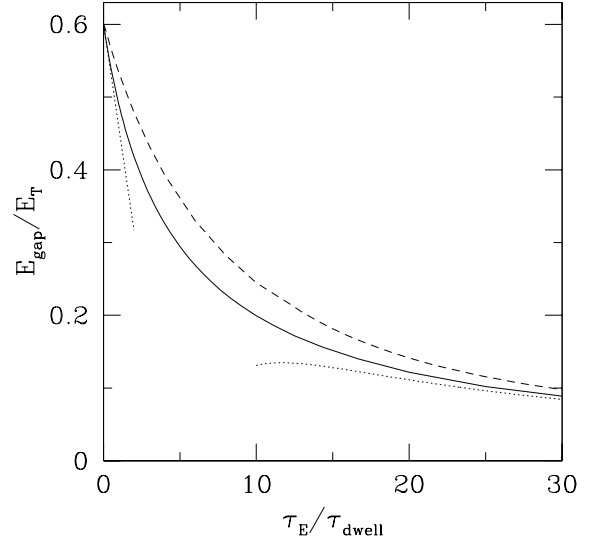}}
\caption{Excitation gap of the Andreev billiard in the crossover from Thouless
to Ehrenfest regimes. The solid curve is the solution of the effective RMT of
Ref.\ \cite{Sil03}, derived in App.\ \ref{gapeffRMT}. The dotted lines are the
two asymptotes (\ref{EgapRMT2}) and (\ref{EgapRMT1}). The dashed curve is the
result of the stochastic model of Ref.\ \cite{Vav03}, discussed in Sec.\
\ref{stochast}. \label{EgaptauE3}
}
\end{figure}

In App.\ \ref{gapeffRMT} we calculate the smallest positive $E$ that solves
Eq.\ (\ref{discreteE2}), which is the excitation gap $E_{\rm gap}$ of the
effective RMT. The result is plotted in Fig.\ \ref{EgaptauE3} (solid curve), as
a function of $\tau_{E}/\tau_{\rm dwell}$. The two asymptotes (dotted lines)
are
\begin{eqnarray}
E_{\rm gap}&=&\frac{\gamma^{5/2}\hbar}{\tau_{\rm
dwell}}\left(1-(2\gamma-1)\frac{\tau_{E}}{\tau_{\rm
dwell}}\right),\;\;\tau_{E}\ll\tau_{\rm dwell},\label{EgapRMT2}\\
E_{\rm gap}&=&\frac{\pi\hbar}{2\tau_{E}}\left(1-(3+\sqrt{8})\frac{\tau_{\rm
dwell}}{\tau_{E}}\right),\;\;\tau_{E}\gg\tau_{\rm dwell},\label{EgapRMT1}
\end{eqnarray}
with $\gamma=\frac{1}{2}(\sqrt{5}-1)$ the golden number.

The $\tau_{E}$ time delay characteristic of the effective RMT was introduced in
Ref.\ \cite{Sil03}, but its effect on the excitation gap was not evaluated
properly.\footnote{I am indebted to P. W. Brouwer for spotting the error.}
As a consequence the formula for the gap given in that paper,
\begin{equation}
E_{\rm gap}=\frac{0.30\,\hbar}{\langle
T\rangle_{\ast}}=\frac{0.30\,\hbar}{\tau_{E}+\tau_{\rm
dwell}},\label{crossover}
\end{equation}
provides only a qualitative description of the actual crossover.

\begin{figure}
\centerline{\includegraphics[width=8cm]{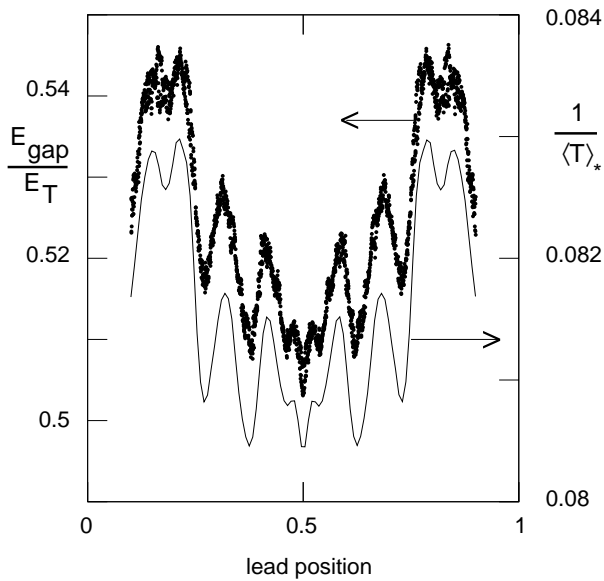}}
\caption{The data points (left axis) are the quantum mechanical gap values
$E_{\rm gap}$ of the Andreev kicked rotator as a function of the location of
the NS interface, for parameter values $M=131072$, $\tau_{\rm dwell}=M/N=5$,
$K=14$. The solid curve (right axis) is the reciprocal of the
mean dwell time (\ref{tastdef}) of classical trajectories longer
than $T^*=7$. Adapted from Ref.\ \cite{Goo03}.
\label{Egap}
}
\end{figure}

The inverse correlation (\ref{crossover}) between the gap and the dwell time of
long trajectories was observed in a computer simulation of the Andreev kicked
rotator \cite{Goo03}. The data points in Fig.\ \ref{Egap} track the excitation
gap as the location in phase space of the NS interface is varied. The solid
curve is a plot of
\begin{equation}
\frac{1}{\langle T \rangle_{\ast}}=
\frac{\int_{T^{\ast}}^{\infty}P(T)dT}{\int_{T^{\ast}}^{\infty}TP(T)dT},
\label{tastdef}
\end{equation}
with $P(T)$ the classical dwell time distribution and $T^{\ast}=7$. We
see that the sample-to-sample fluctuations in the gap correlate very
well with the fluctuations in the mean dwell time
of long trajectories. The correlation is not sensitive to the choice
of $T^{\ast}$, as long as it is greater than $\tau_{E}=4.4$.

\subsection{Stochastic model}
\label{stochast}

Small-angle scattering by a smooth disorder potential provides a stochastic
model for the quantum diffraction of a wave packet in a chaotic billiard
\cite{Ale96}. The scattering time of the stochastic model plays the role of the
Ehrenfest time in the deterministic chaotic dynamics. The advantage of a
stochastic description is that one can average over different realizations of
the disorder potential. This provides for an established set of analytical
techniques. The disadvantage is that one does not know how well stochastic
scattering mimics quantum diffraction.

Vavilov and Larkin \cite{Vav03} have used the stochastic model to study the
crossover from the Thouless regime to the Ehrenfest regime in an Andreev
billiard. They discovered that the rapid turn-on of quantum diffraction at
$\tau_{E}\agt\tau_{\rm dwell}$ not only causes an excitation gap to open at
$\hbar/\tau_{E}$, but that it also causes oscillations with period
$\hbar/\tau_{E}$ in the ensemble-averaged density of states
$\langle\rho(E)\rangle$ at high energies $E\agt E_{T}$. In normal billiards
oscillations with this periodicity appear in the level-level correlation
function \cite{Ale97}, but not in the level density itself.

\begin{figure}
\centerline{\includegraphics[width=8cm]{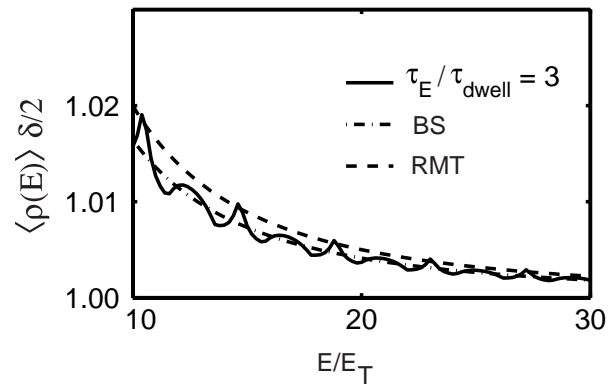}}
\caption{Oscillatory density of states at finite Ehrenfest time (solid curve),
compared with the smooth limits of zero (RMT) and infinite (BS) Ehrenfest
times. The solid curve is the result of the stochastic model of Vavilov and
Larkin, for $\tau_{E}=3\,\tau_{\rm dwell}=3\hbar/2E_{T}$. (The definition
(\ref{tauEdef}) of the Ehrenfest time used here differs by a factor of two from
that used by those authors.) Adapted from Ref.\ \cite{Vav03}.
\label{oscillations}
}
\end{figure}

The predicted oscillatory high-energy tail of $\langle\rho(E)\rangle$ is
plotted in Fig.\ \ref{oscillations}, for the case $\tau_{E}/\tau_{\rm
dwell}=3$, together with the smooth results of RMT ($\tau_{E}/\tau_{\rm
dwell}\rightarrow 0$) and Bohr-Sommerfeld (BS) theory ($\tau_{E}/\tau_{\rm
dwell}\rightarrow\infty$).

Independent analytical support for the existence of oscillations in the density
of states with period $\hbar/\tau_{E}$ comes from the singular perturbation
theory of Ref.\ \cite{Ada02b}. Support from numerical simulations is still
lacking. Jacquod et al.\ \cite{Jac03} did find pronounced oscillations for
$E\agt E_{T}$ in the level density of the Andreev kicked rotator. However,
since these could be described by the Bohr-Sommerfeld theory they can not be
the result of quantum diffraction, but must be due to nonergodic trajectories
\cite{Ihr01b}.

The $\tau_{E}$-dependence of the gap obtained by Vavilov and Larkin is plotted
in Fig.\ \ref{EgaptauE3} (dashed curve). It is close to the result of the
effective RMT (solid curve). The two theories predict the same limit $E_{\rm
gap}\rightarrow\pi\hbar/2\tau_{E}$ for $\tau_{E}/\tau_{\rm
dwell}\rightarrow\infty$. The asymptotes given in Ref.\ \cite{Vav03} are
\begin{eqnarray}
E_{\rm gap}&=&\frac{\gamma^{5/2}\hbar}{\tau_{\rm
dwell}}\left(1-0.23\,\frac{\tau_{E}}{2\tau_{\rm
dwell}}\right),\;\;\tau_{E}\ll\tau_{\rm dwell},\label{EgapRMT2b}\\
E_{\rm gap}&=&\frac{\pi\hbar}{2\tau_{E}}\left(1-\frac{2\tau_{\rm
dwell}}{\tau_{E}}\right),\;\;\tau_{E}\gg\tau_{\rm dwell}.\label{EgapRMT1b}
\end{eqnarray}
Both are different from the results (\ref{EgapRMT2}) and (\ref{EgapRMT1}) of
the effective RMT.\footnote{
Since $2\gamma-1=0.236$, the small-$\tau_{E}$ asymptote of Vavilov and Larkin
differs by a factor of two from that of the effective RMT.}

\subsection{Numerical simulations}
\label{numerics}

\begin{figure}
\centerline{\includegraphics[width=8cm]{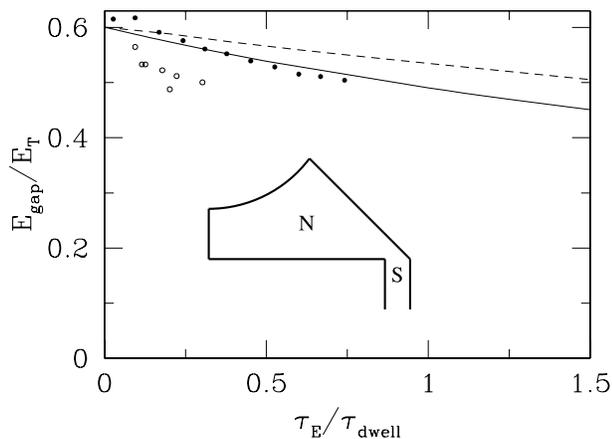}}
\caption{Ehrenfest-time dependence of the excitation gap in an Andreev
billiard, according to the effective RMT (solid curve, calculated in App.\
\ref{gapeffRMT}) and according to the stochastic model (dashed curve,
calculated in Ref.\ \cite{Vav03}). The data points result from the simulation
of the Andreev kicked rotator \cite{Jac03} (closed circles, in the range
$N=10^{2}-10^{5}$) and of the Sinai billiard shown in the inset \cite{Kor04}
(open circles, in the range $N=18-30$).
\label{comparison2}
}
\end{figure}

Because the Ehrenfest time grows only logarithmically with the size of the
system, it is exceedingly difficult to do numerical simulations deep in the
Ehrenfest regime. Two simulations \cite{Jac03,Kor04} have been able to probe
the initial decay of the excitation gap, when $\tau_{E}\alt\tau_{\rm dwell}$.
We show the results of both simulations in Fig.\ \ref{comparison2} (closed and
open circles), together with the full decay as predicted by the effective RMT
of Sec.\ \ref{EffRMT} (solid curve) and by the stochastic model of Sec.\
\ref{stochast} (dashed curve).

The closed circles were obtained by Jacquod et al.\ \cite{Jac03} using the
stroboscopic model of Sec.\ \ref{strobomodel} (the Andreev kicked rotator). The
number of modes $N$ in the contact to the superconductor was increased from
$10^2$ to $10^5$ at fixed dwell time $\tau_{\rm dwell}=M/N=5$ and kicking
strength $K=14$ (corresponding to a Lyapunov exponent
$\alpha\approx\ln(K/2)=1.95$). In this way all classical properties of the
billiard remain the same while the effective Planck constant $h_{\rm
eff}=1/M=1/N\tau_{\rm dwell}$ is reduced by three orders of magnitude. To plot
the data as a function of $\tau_{E}/\tau_{\rm dwell}$, Eq.\ (\ref{tauEdef}) was
used for the Ehrenfest time. The unspecified terms of order unity in that
equation were treated as a single fit parameter. (This amounts to a horizontal
shift by $-0.286$ of the data points in Fig.\ \ref{comparison2}.)

The open circles were obtained by Korm\'{a}nyos et al.\ \cite{Kor04} for the
chaotic Sinai billiard shown in the inset. The number of modes $N$ was varied
from 18 to 30 by varying the width of the contact to the superconductor. The
Lyapunov exponent $\alpha\approx 1.7$ was fixed, but $\tau_{\rm dwell}$ was not
kept constant in this simulation. The Ehrenfest time was computed by means of
the same formula (\ref{tauEdef}), with $M=2L_{c}k_{F}/\pi$ and $L_{c}$ the
average length of a trajectory between two consecutive bounces at the curved
boundary segment.

The data points from both simulations have substantial error bars (up to 10\%).
Because of that and because of their limited range, we can not conclude that
the simulations clearly favor one theory over the other.

\section{Conclusion}
\label{concl}

Looking back at what we have learned from the study of Andreev billiards, we
would single out the breakdown of random-matrix theory as the most unexpected
discovery and the one with the most far-reaching implications for the field of
quantum chaos. In an isolated chaotic billiard RMT provides an accurate
description of the spectral statistics on energy scales below $\hbar/\tau_{\rm
erg}$ (the inverse ergodic time). The weak coupling to a superconductor causes
RMT to fail at a much smaller energy scale of $\hbar/\tau_{\rm dwell}$ (the
inverse of the mean time between Andreev reflections), once the Ehrenfest time
$\tau_{E}$ becomes greater than $\tau_{\rm dwell}$.

In the limit $\tau_{E}\rightarrow\infty$, the quasiclassical Bohr-Sommerfeld
theory takes over from RMT. While in isolated billiards such an approach can
only be used for integrable dynamics, the Bohr-Sommerfeld theory of Andreev
billiards applies regardless of whether the classical motion is integrable or
chaotic. This is a demonstration of how the time-reversing property of Andreev
reflection unravels chaotic dynamics.

What is lacking is a conclusive theory for finite $\tau_{E}\agt\tau_{\rm
dwell}$. The two phenomenological approaches of Secs.\ \ref{EffRMT} and
\ref{stochast} agree on the asymptotic behavior
\begin{equation}
\lim_{\hbar\rightarrow 0}E_{\rm gap}=\frac{\pi\hbar\alpha}{2|\ln \hbar|+{\rm
constant}},
\end{equation}
in the classical $\hbar\rightarrow 0$ limit (understood as $N\rightarrow\infty$
at fixed $\tau_{\rm dwell}$). There is still some disagreement on how this
limit is approached. We would hope that a fully microscopic approach, for
example based on the ballistic $\sigma$-model \cite{Muz95,And96}, could provide
a conclusive answer. At present technical difficulties still stand in the way
of a solution along those lines \cite{Tar01}.

A new direction of research is to investigate the effects of a nonisotropic
superconducting order parameter on the Andreev billiard. The case of $d$-wave
symmetry is most interesting because of its relevance for high-temperature
superconductors. The key ingredients needed for a theoretical description
exist, notably RMT \cite{Alt02}, quasiclassics \cite{Ada02c}, and a numerically
efficient Andreev map \cite{Ada04}.

\section*{Acknowledgments}
While writing this review, I benefitted from correspondence and discussions
with W. Belzig, P. W. Brouwer, J. Cserti, P. M. Ostrovsky, P. G. Silvestrov,
and M. G. Vavilov. The work was supported by the Dutch Science Foundation
NWO/FOM.

\appendix
\section{Excitation gap in effective RMT and relationship with delay times}
\label{gapeffRMT}

We seek the edge of the excitation spectrum as it follows from the determinant
equation (\ref{discreteE2}), which in zero magnetic field and for $E\ll\Delta$
takes the form
\begin{equation}
{\rm
Det}\,\left[1+e^{2iE\tau_{E}/\hbar}S_{0}(E)S_{0}(-E)^{\dagger}\right]=0.
\label{discreteE3}
\end{equation}
The unitary symmetric matrix $S_{0}$ has the RMT distribution of a chaotic
cavity with effective parameters $N_{\rm eff}$ and $\delta_{\rm eff}$ given by
Eqs.\ (\ref{Neffdef}) and (\ref{deltaeffdef}). The mean dwell time associated
with $S_{0}$ is $\tau_{\rm dwell}$. The calculation for $N_{\rm eff}\gg 1$
follows the method described in Secs.\ \ref{RMTeffH} and \ref{RMTexcgap},
modified as in Ref.\ \cite{Bro97a} to account for the energy dependent phase
factor in the determinant.

Since $S_{0}$ is of the RMT form (\ref{SWHformula}), we can write Eq.\
(\ref{discreteE3}) in the Hamiltonian form (\ref{DetEHW}). The extra phase
factor $\exp(2iE\tau_{E}/\hbar)$ introduces an energy dependence of the
coupling matrix,
\begin{equation}
{\cal W}_{0}(E)=
\frac{\pi}{\cos u}
\left(\begin{array}{cc}
W_{0}W_{0}^{T}\sin u&W_{0}W_{0}^{T}\\
W_{0}W_{0}^{T}&W_{0}W_{0}^{T}\sin u
\end{array}\right),\label{calWdefnew}
\end{equation}
where we have abbreviated $u=E\tau_{E}/\hbar$. The subscript $0$ reminds us
that the coupling matrix refers to the reduced set of $N_{\rm eff}$ channels in
the effective RMT. Since there is no tunnel barrier in this case, the matrix
$W_{0}$ is determined by Eq.\ (\ref{Wgammarelation}) with $\Gamma_{n}\equiv 1$.
The Hamiltonian
\begin{equation}
{\cal H}_{0}=\left(\begin{array}{cc}
H_{0}&0\\0&-H_{0}
\end{array}\right)\label{calHdefnew}
\end{equation}
is that of a chaotic cavity with mean level spacing $\delta_{\rm eff}$. We seek
the gap in the density of states
\begin{equation}
\rho(E)=-\frac{1}{\pi}{\rm Im}\,{\rm Tr}\,\left(1+\frac{d{\cal
W}_{0}}{dE}\right)(E + i0^{+}-{\cal H}_{0}+{\cal
W}_{0})^{-1},\label{resolventnew}
\end{equation}
cf.\ Eq.\ (\ref{resolvent}).

The selfconsistency equation for the ensemble-averaged Green function,
\begin{equation}
{\cal G}=[E+{\cal W}_{0}-(M\delta_{\rm
eff}/\pi)\sigma_{z}G\sigma_{z}]^{-1},\label{pastur1new}
\end{equation}
still leads to Eq.\ (\ref{dysongamma1}), but Eq.\ (\ref{dysongamma2}) should be
replaced by
\begin{eqnarray}
G_{11}+G_{12}\sin u&=&-(\tau_{\rm dwell}/\tau_{E})uG_{12}\nonumber\\
&&\mbox{}\times(G_{12}+\cos u+G_{11}\sin u).\label{dysongamma2new}
\end{eqnarray}
(We have used that $N_{\rm eff}\delta_{\rm eff}=2\pi\hbar/\tau_{\rm dwell}$.)
Because of the energy dependence of the coupling matrix, the equation
(\ref{onemoretrace}) for the ensemble averaged density of states should be
replaced by
\begin{equation}
\langle\rho(E)\rangle=-\frac{2}{\delta}\,{\rm Im}\,\left(G_{11}-\frac{u}{\cos
u}G_{12}\right).\label{onemoretracenew}
\end{equation}

The excitation gap corresponds to a square root singularity in
$\langle\rho(E)\rangle$, which can be obtained by solving Eqs.\
(\ref{dysongamma1}) and (\ref{dysongamma2new}) jointly with $dE/dG_{11}=0$ for
$u\in(0,\pi/2)$. The result is plotted in Fig.\ \ref{EgaptauE3}. The small- and
large-$\tau_{E}$ asymptotes are given by Eqs.\ (\ref{EgapRMT2}) and
(\ref{EgapRMT1}).

The large-$\tau_{E}$ asymptote is determined by the largest eigenvalue of the
time-delay matrix. To see this relationship, note that for
$\tau_{E}\gg\tau_{\rm dwell}$ we may replace the determinant equation
(\ref{discreteE3}) by
\begin{equation}
{\rm Det}\,\left[1+\exp[2iE\tau_{E}/\hbar+2iEQ(0)]+{\cal
O}\left(\frac{\tau_{\rm
dwell}}{\tau_{E}}\right)^{2}\right]=0.\label{discreteE4}
\end{equation}
The Hermitian matrix
\begin{equation}
Q(E)=\frac{1}{i}S_{0}(E)^{\dagger}\frac{d}{dE}S_{0}(E)\label{Qdef}
\end{equation}
is known in RMT as the Wigner-Smith or time-delay matrix.
The roots $E_{np}$ of Eq.\ (\ref{discreteE4}) satisfy
\begin{equation}
2E_{np}(\tau_{E}+\tau_{n})=(2p+1)\pi\hbar,\;\;p=0,1,2,\ldots.\label{rooteq}
\end{equation}
The eigenvalues $\tau_{1},\tau_{2},\ldots\tau_{N_{\rm eff}}$ of $\hbar Q(0)$
are the delay times. They are all positive, distributed according to a
generalized Laguerre ensemble \cite{Bro98}. In the limit $N_{\rm
eff}\rightarrow\infty$ the distribution of the $\tau_{n}$'s is nonzero only in
the interval $(\tau_{-},\tau_{+})$, with $\tau_{\pm}=\tau_{\rm
dwell}(3\pm\sqrt{8})$. By taking $p=0$, $\tau_{n}=\tau_{+}$ we arrive at the
asymptote (\ref{EgapRMT1}).

The precise one-to-one correspondence (\ref{rooteq}) between the spectrum of
low-lying energy levels of the Andreev billiard and the spectrum of delay times
is a special property of the classical limit $\tau_{E}\rightarrow\infty$. For
$\tau_{E}\alt\tau_{\rm dwell}$ there is only a qualitative similarity of the
two spectral densities \cite{Cra02}.

\end{document}